\documentclass[aps,pre,showpacs,superscriptaddress,groupedaddress,twocolumn]{revtex4}
\usepackage{graphicx}  
\usepackage{bm}        
\usepackage{amssymb}   
\usepackage{amsmath}
\usepackage{enumerate}
\usepackage{color}
\usepackage{pgfplots}
\usepackage{tikz}
\usetikzlibrary{calc,decorations.markings}
\begin{document}

\title{Stochastic thermodynamics of Langevin systems under time-delayed feedback control: II. Nonequilibrium  steady-state fluctuations}
\author{ M.L. Rosinberg}
\affiliation{Laboratoire de Physique Th\'eorique de la Mati\`ere Condens\'ee, Universit\'e Pierre et Marie Curie,CNRS UMR 7600,\\ 4 place Jussieu, 75252 Paris Cedex 05, France}
\email{mlr@lptmc.jussieu.fr}
\author{G. Tarjus}
\affiliation{Laboratoire de Physique Th\'eorique de la Mati\`ere Condens\'ee, Universit\'e Pierre et Marie Curie,CNRS UMR 7600,\\ 4 place Jussieu, 75252 Paris Cedex 05, France}
\email{tarjus@lptmc.jussieu.fr}\author{T. Munakata}
\affiliation{Department of Applied Mathematics and Physics, 
Graduate School of Informatics, Kyoto University, Kyoto 606-8501, Japan
} 

\begin{abstract}
This paper is the second in a series devoted to the study of Langevin systems subjected to a continuous time-delayed feedback control. The goal of our previous paper [Phys. Rev. E {\bf 91}, 042114 (2015)]  was to derive second-law-like inequalities that provide bounds to the average extracted work.  Here we study stochastic fluctuations of time-integrated observables such as the  heat exchanged with the environment, the extracted work, or the (apparent) entropy production. We use a path-integral formalism and focus on the long-time behavior in the stationary cooling regime, stressing the role of rare events. This is illustrated by a detailed analytical and numerical study  of  a Langevin harmonic oscillator driven by a linear feedback. 

\end{abstract} 

 
\maketitle

\section{Introduction}

This paper is part of an ongoing effort to include the effect of time delay in the thermodynamic description of small stochastic systems subjected to a continuous feedback control. Time delay is now recognized to play an essential role in many physical, biological, and information systems and also occurs very frequently in experimental setups. Moreover, within the last two decades, including a delay between the detection and the control operation has emerged as an important feedback strategy for  controlling transport  or stabilizing irregular motion in classical or quantum systems, especially in the presence of noise (see e.g. the collection of papers in \cite{A2010,JPSS2010,SKH2016}). Accordingly, there is much interest in the mathematical and control theory literature for exploring the plethora of complex phenomena produced by the combination of time delay and noise.

On the other hand, it is much less common to analyze time-delayed feedback loops from the perspective of energetic  and information exchanges, which is the main focus of the emerging fields of stochastic and information thermodynamics~\cite{S2012,PHS2015}. One reason is the non-Markovian nature of the dynamics which makes the theoretical description more challenging (for instance, one cannot resort to a spectral approach using Fokker-Planck operators). 
This is not an impossible task, though, and in a previous work~\cite{RMT2015}, hereafter referred to as I, we have initiated a theoretical study of an underdamped  Langevin equation that models the motion of a nanomechanical resonator in contact  with a thermal reservoir and subjected to a time-delayed, position-dependent force. The role of the control force is to damp thermal fluctuations  and to maintain the resonator in a nonequilibrium steady state (NESS) where its average (configurational or kinetic) temperature is much smaller than  the temperature of the environment. Heat is thus permanently extracted from the bath and converted into work, which means that the feedback control operates as an autonomous Maxwell's demon. We then derived  a series of second-law-like inequalities that provide bounds to the average extracted work. One of these bounds, obtained by (formally) time reversing the feedback, is intimately related to the non-Markovian character of the dynamics.

However, fluctuations dominate at the nanoscale~\cite{J2011}, and it is not sufficient to merely describe  observables by their typical value. It is also important to study the large-deviation statistics that characterizes the fluctuations  at long times. 
This is the purpose of the present work where we extend the study of I by considering the nonequilibrium fluctuations  of three time-integrated thermodynamic quantities, the heat, the work, and a so-called ``apparent" entropy production (to be defined below). These observables have the same average value in the stationary state but their fluctuations may differ because of the unbounded growth of temporal boundary terms. As it turns out, these fluctuations are very dependent on the time delay, and this issue is the central theme of this work.  

The paper is organized as follows. We first review in section II  some basic  facts  about the model, the observables, and the calculation of the large deviation rate functions.  Then, in section III, we introduce two different conjugate dynamics and use them to derive two expressions of the dissipated heat as a ratio of path probabilities.  This allows us to express the  path-integral representations of  the cumulant generating functions in three different ways, which will play an important role in our study.   Section IV, which is the central part of the paper, is devoted to a detailed numerical and analytical study of the large-deviation statistics for a harmonic oscillator driven by a linear feedback. The main objective of the theoretical analysis is to explain the intriguing effect of the delay on the probability distributions of the observables in the long-time limit. Special attention is paid to the behavior of the corresponding scaled cumulant generating functions and to the connection between rare fluctuations  of the temporal boundary terms and the asymptotic behavior of the conjugate dynamics. We finally derive two stationary-state fluctuation theorems for the  work performed by the feedback force.  Summary and closing remarks are presented in Section V.  Some additional but important pieces of information are given in two Appendices. In particular, Appendix B offers a complete analytical study of the fluctuations in the (Markovian) small-delay limit where the feedback generates an additional viscous damping and the so-called ``molecular refrigerator" model studied in~\cite{KQ2004,KQ2007,MR2012} is recovered.

We have tried to make the present paper self-consistent as much as possible. However, we warn the reader that some analytical developments relies strongly on paper I, in particular on Section V.B.2.

\section{Model and observables}

As in paper I, we consider an underdamped Brownian particle of mass $m$ immersed in a thermal environment with viscous damping $\gamma$ and temperature $T$. The dynamical  evolution is governed  by the one-dimensional  Langevin equation 
\begin{align}
\label{EqL1}
m\dot  v_t=-\gamma v_t+F(x_t)+F_{fb}(t)+\sqrt{2\gamma T}\xi_t
\end{align}
where $v_t=\dot x_t$, $F(x)=-dV(x)/dx$ is a conservative force, and $\xi_t$ is a zero-mean Gaussian white noise with unit variance (throughout the paper, temperatures and entropies are measured in units of the Boltzmann constant $k_B$). $F_{fb}(t)$ is the  feedback control force which  depends on the position of the particle at time $t-\tau$:
\begin{align}
F_{fb}(t)= F_{fb}(x_{t-\tau}) \ ,
\end{align}
where $\tau>0$ is the time delay. This model is intended to describe an autonomous feedback process in which the instantaneous state of the system (here, the position of the Brownian particle) is continuously monitored with perfect accuracy, but some time is needed to implement the control.  Clearly, $\tau$ must be  smaller than any relaxation time in the system for the control to be efficient. We stress that it is the stochastic force $F_{fb}(t)$ that makes the system's dynamics non-Markovian and not the interaction with the environment. 

Our goal  is to study the fluctuations of a  time-integrated observable ${\cal A}_t$ such as the work done by the feedback force  or the heat exchanged with the environment during the time interval $[0,t]$, assuming that the system has reached a nonequilibrium steady state (NESS). As discussed in I, this requires to properly choose the parameters of the feedback loop, such as the delay or the feedback gain.  In fact, multiple NESS may exist, which is a remarkable feature of time-delayed systems (see e.g. Fig \ref{Fig2} below). Moreover, we will focus on regions of the parameter space where the feedback controller acts as a Maxwell's demon who permanently extracts heat from the environment and uses it  as work to maintain the system at a  temperature smaller than $T$.

The time-integrated work and  dissipated heat are defined as
\begin{align}
\label{EqWXY}
{\cal W}_t[{\bf X},{\bf Y}]=\int_0^t dt' \: F_{fb}(x_{t'-\tau})\circ v_{t'} \ ,
\end{align}
and
\begin{align}
\label{EqQXY}
{\cal Q}_t[{\bf X},{\bf Y}]&=\int_0^t dt'\: \Big[\gamma v_{t'}-\sqrt{2\gamma T}\xi(t)] \circ v_{t'}\nonumber\\
&= -\int_0^t dt'\: \Big[m \dot v_{t'} -F(x_{t'})-F_{fb}(x_{t'-\tau})\Big] \circ v_{t'} \ ,
\end{align}
where the integrals are interpreted with the Stratonovich prescription. These are standard definitions of work and heat in stochastic thermodynamics~\cite{Sbook2010,S2012}, except for the fact that the delay makes the two observables depending on both ${\bf X}$,  the system trajectory in phase space in the time interval $[0,t]$, and  ${\bf Y}$, the trajectory  in the previous interval $[-\tau,0]$ (we  here assume that $t\ge\tau$ so that ${\bf x}_i\equiv (x_0,v_0)\equiv{\bf y}_f$).  This of course is a source of complication for the theoretical description, although one may suspect that the dependence on ${\bf Y}$ does not play a major role at long times. From now on, we  will drop the functional dependence of the observables on  ${\bf X}$ and ${\bf Y}$ to simplify the notation. (There are a few other differences with the notations used in I: the time window is now $[0,t]$ instead of $[-{\cal T}, {\cal T}]$ and the time-integrated observables are denoted by calligraphic uppercase symbols, e.g. ${\cal W}_t$ instead of $w$.)

In the following, we will also consider the fluctuations of the trajectory-dependent functional (dubbed as an ``apparent" entropy production)
\begin{align}
\label{EqSigmaXY}
\Sigma_t= \Sigma^m_t +\ln \frac{p_0({\bf x}_i)}{p_1({\bf x}_f)}\ ,
\end{align}
where $\Sigma^m_t=\beta {\cal Q}_t$  ($\beta =(k_BT)^{-1}$)  is the entropy change in the medium,  ${\bf x}_f\equiv (x_t,v_t)$, and $p_0({\bf x}), p_1({\bf x})$ are arbitrary normalized distributions. In the steady state, the natural choice for these distributions is $p_0({\bf x})=p_1({\bf x})=p_{st}({\bf x})$, and an observer unaware of the existence of the feedback control would  regard $\Sigma_t$ as the total stochastic entropy production (EP) in the time interval $[0,t]$~\cite{S2005}. However, $\Sigma_t$ is negative on average in the cooling regime, in apparent violation of the second law, and more generally  does not obey a fluctuation theorem, $\langle e^{-\Sigma_t}\rangle\ne 1$.  Another, but more complicated, trajectory-dependent functional that may quantify the entropy production in the system was introduced in I. This functional does satisfy an IFT.

It is important to notice that the three fluctuating quantities $\beta {\cal W}_t,\beta {\cal Q}_t$, and $\Sigma_t$  have the same expectation value in the stationary state
\begin{align}
\langle{\beta\cal W}_t\rangle_{st}=\langle {\beta \cal Q}_t\rangle_{st}=\langle \Sigma_t\rangle_{st}\ .
\end{align}
Moreover, ${\cal W}_t$ and ${\cal Q}_t$ are related via the first law that expresses the conservation of energy at the microscopic level~\cite{Sbook2010},
\begin{align}
\label{Eq1law}
{\cal Q}_t={\cal W}_t-\Delta {\cal U}({\bf x}_i,{\bf x}_f)\ ,
\end{align}
where 
\begin{align}
\Delta {\cal U}({\bf x}_i,{\bf x}_f)=\frac{1}{2}m(v_t^2-v_0^2)+V(x_t)-V(x_0) 
\end{align}
is the change in the internal energy of the system after the time $t$. Accordingly, the fluctuations of ${\cal W}_t$, ${\cal Q}_t$, and $\Delta {\cal U}$ are not independent.

We are interested in the long-time behavior of the stationary probability distribution functions (pdfs)  $P_{st}({\cal A}_t)$, where ${\cal A}_t$ stands for either $\beta{\cal W}_t$ or $\beta{\cal Q}_t$ or $\Sigma_t$. As $t \to \infty$,  we expect these pdfs to acquire the scaling form
\begin{align}
P_{st}({\cal A}_t=at)\sim e^{-I(a)t}
\end{align}
where $I(a)\equiv -\lim_{t\to \infty} (1/t)\ln P_{st}({\cal A}_t=at)$ is the large deviation rate  function (LDF) that is used to characterize the statistics of exponentially rare events~\cite{T2009}.  As usual, to obtain the rate function, we introduce the moment generating or characteristic function 
\begin{align}
\label{EqZ}
Z_A(\lambda,t)&=\langle e^{-\lambda{\cal A}_t}\rangle_{st}
\end{align}
and the corresponding scaled cumulant generating function (SCGF)
\begin{align}
\label{EqmuAlambda}
\mu_A(\lambda)\equiv\lim_{{t}\rightarrow \infty}\frac{1}{t} \ln \langle e^{-\lambda {\cal A}_t}\rangle_{st}
\end{align}
whose  behavior  away from $\lambda =0$ encodes information about rare trajectories contributing to the tails of the pdf. For generic values of $\lambda$, one expects $\mu_A(\lambda)$ to be the same function $\mu(\lambda)$ for $\beta{\cal W}_t,\beta{\cal Q}_t$, and $\Sigma_t$ since the three observables only differ by temporal boundary terms like $\Delta {\cal U}({\bf x}_i,{\bf x}_f)$ 
or $\ln p_0({\bf x}_i)/p_1({\bf x}_f)$. This amounts to assuming that the generating functions behave asymptotically as
\begin{align}
\label{EqZ0}
Z_A(\lambda,t)\sim g_A(\lambda)e^{\mu(\lambda)t}\ ,
\end{align}
where the dependence on the observable ${\cal A}_t$ is included in the subleading factor $g_A(\lambda)$ that results from the average over the initial and final states (in the present case, the initial ``state" involves the whole trajectory ${\bf Y}$).  The LDFs $I(a)$ are then obtained via the Legendre transform 
\begin{align}
\label{EqLT}
I(a)=-\lambda^* a-\mu(\lambda^*)\ ,
\end{align}
with the saddle point $\lambda^*(a)$ being the root of $\mu'(\lambda^*)=-a$~\cite{T2009}. However,  Eq. (\ref{EqLT}) breaks down when $g_A(\lambda)$  has  singularities in the region of the saddle-point integration due to rare but large fluctuations of the boundary terms.  Although such terms typically do not grow with time, they may indeed fluctuate to order $t$ when the potential $V(x)$ is unbounded, which is the situation considered here. The leading contribution to the LDF then comes from the singularity, which induces  an exponential tail in the pdf. This issue is now well documented in the literature, both theoretically~\cite{VC2003,Fa2002,V2006,BJMS2006,PRV2006,HRS2006,TC2007,Sab2011,N2012,NP2012,KKP2014} and experimentally~\cite{GC2005,F2008,B2009}. In consequence, while the three observables $\beta{\cal W}_t,\beta{\cal Q}_t$, and $\Sigma_t$ have the same expectation value, their LDFs  may differ.
 In some circumstances, large fluctuations of the boundary terms  may even induce a discontinuity of the SCGF at $\lambda=1$,  as pointed out recently~\cite{RTM2016}: the asymptotic expression  (\ref{EqZ0}) is then no longer valid and $\mu_A(1)\ne \mu(1)$. We shall see later on that this is very much dependent on the time delay.
 
The main difficulty we are facing in the present study is that no analytical methods are currently known to compute the SCGFs. If the dynamics were Markovian, one would determine the largest eigenvalue of the appropriate Fokker-Planck operator ~\cite{T2009}. But there is no such operator in the presence of delay (except in the small-$\tau$ limit where Markovianity is recovered), and one has to rely on numerical simulations or to focus on a linear dynamics for which the calculation of $\mu(\lambda)$ can be carried out by going to the frequency domain.
However, even in this case, the expression of the prefactors $g_A(\lambda)$ remains out of reach for generic values of $\lambda$.

\section{Conjugate dynamics and generating functions}

\subsection {Conjugate dynamics and dissipated heat}

We begin our study by recalling two expressions for the heat dissipated along a trajectory that will play a significant role in the following. We stress that these relations are valid for trajectories of arbitrary duration. There is no need to take the limit $t\to \infty$.

The first relation is obtained by introducing a modified or ``conjugate" Langevin dynamics in which the sign of the viscous damping is flipped, i.e.,
\begin{align}
\label{EqL1minus}
m\dot  v_t=\gamma v_t+F(x_t)+F_{bf}(x_{t-\tau})+\sqrt{2\gamma T}\xi_t\ .
\end{align}
This readily yields~\cite{RTM2016}
\begin{align}
\label{EqELB2}
\beta  {\cal Q}_t=\ln \frac{{\cal P}[ {\bf X}\vert {\bf Y}]}{\hat{\cal P}[{\bf X}\vert {\bf Y}]}-\frac{\gamma}{m}t \ ,
\end{align}
where ${\cal P}[{\bf X}\vert {\bf Y}]$ and $\hat {\cal P}[ {\bf X}\vert {\bf Y}]$ are the  conditional probabilities of realizing the trajectory ${\bf X}$ with the original and conjugate dynamics, respectively,  given the trajectory ${\bf Y}$ (and thus the initial value ${\bf x}_i={\bf y}_f$). These two probabilities can be expressed in terms of Onsager-Machlup (OM) action functionals~\cite{OM1953},
\begin{subequations}
\label{Path:subeqns}
\begin{align}
{\cal P}[{\bf X}\vert{\bf Y}]&\propto e^{\frac{\gamma}{2m}t}  \: e^{-\beta  {\cal S}[{\bf X},{\bf Y}]} \label{Path:subeq1}\\
\hat{\cal P}[ {\bf X}\vert {\bf Y}]&\propto e^{-\frac{\gamma}{2m}t}  \: e^{-\beta  \hat{\cal S}[{\bf X},{\bf Y}]} \label{Path:subeq2}\ ,
\end{align}
\end{subequations} 
where
\begin{subequations}
\label{Actions:subeqns}
\begin{align}
 {\cal S}[{\bf  X},{\bf Y}]&=\frac{1}{4\gamma}\int_{0}^{t}dt' \:\Big[m\dot v_{t'}+\gamma v_{t'}-F(x_{t'})-F_{fb}(x_{t'-\tau})\Big]^2 \label{Actions:subeq1}\\
\hat{\cal S}[{\bf X},{\bf Y}]&=\frac{1}{4\gamma}\int_{0}^{t}dt' \:\Big[m\dot v_{t'}-\gamma v_{t'}-F(x_{t'})-F_{fb}(x_{t'-\tau})\Big]^2 \label{Actions:subeq2} \ ,
\end{align}
\end{subequations}
and the exponential factors $e^{\pm\frac{\gamma }{2m}t}$ come from the Jacobians  of the transformations $\xi(t) \rightarrow x(t)$ associated with the two Langevin dynamics (see ~\cite{IP2006} or the supplemental material of ~\cite{RTM2016} for a derivation).   As usual, the continuous-time integrals in Eqs. (\ref{Actions:subeqns}) are interpreted as the limit of discrete sums, as discussed for instance in the Appendix B of ~\cite{ABC2010}. We recall that there is no need to specify the interpretation  (Ito versus Stratonovitch) of the stochastic calculus as long as $m\ne 0$.  From now on, the {\it hat} symbol   will  refer to quantities associated with the  $\gamma \to -\gamma$ conjugate dynamics (\ref{EqL1minus}).

 From Eq. (\ref{EqELB2}), one immediately obtains an integral fluctuation theorem (IFT) for the dissipated heat~\cite{RTM2016},
 \begin{align}
\label{EqIFTQ}
\langle e^{-\beta  {\cal Q}_t}\rangle=e^{\frac{\gamma}{m}t}\ .
\end{align} 
In particular, this implies at long times that 
 \begin{align}
\label{EqmuQ1}
\mu_Q(1)=\frac{\gamma}{m}\ .
\end{align}
Moreover, the average dissipated heat satisfies $\langle \beta  {\cal Q}_t\rangle\ge -(\gamma/m)t$ by Jensen's inequality.  This  bound is trivial, though, and can be directly obtained by averaging Eq. (\ref{EqQXY}), which yields \begin{align}
\label{EqQav}
\langle {\cal Q}_t\rangle=\frac{\gamma}{m}\int_0^t dt' (T_v(t')-T)\ ,
\end{align}
where $T_v(t)=m\langle v^2_{t}\rangle\ge 0$ is the effective temperature of the momentum degree of freedom. 

\begin{figure}[hbt]
\begin{center}
\includegraphics[trim={0cm 0cm 0cm 0cm},clip,width=6cm]{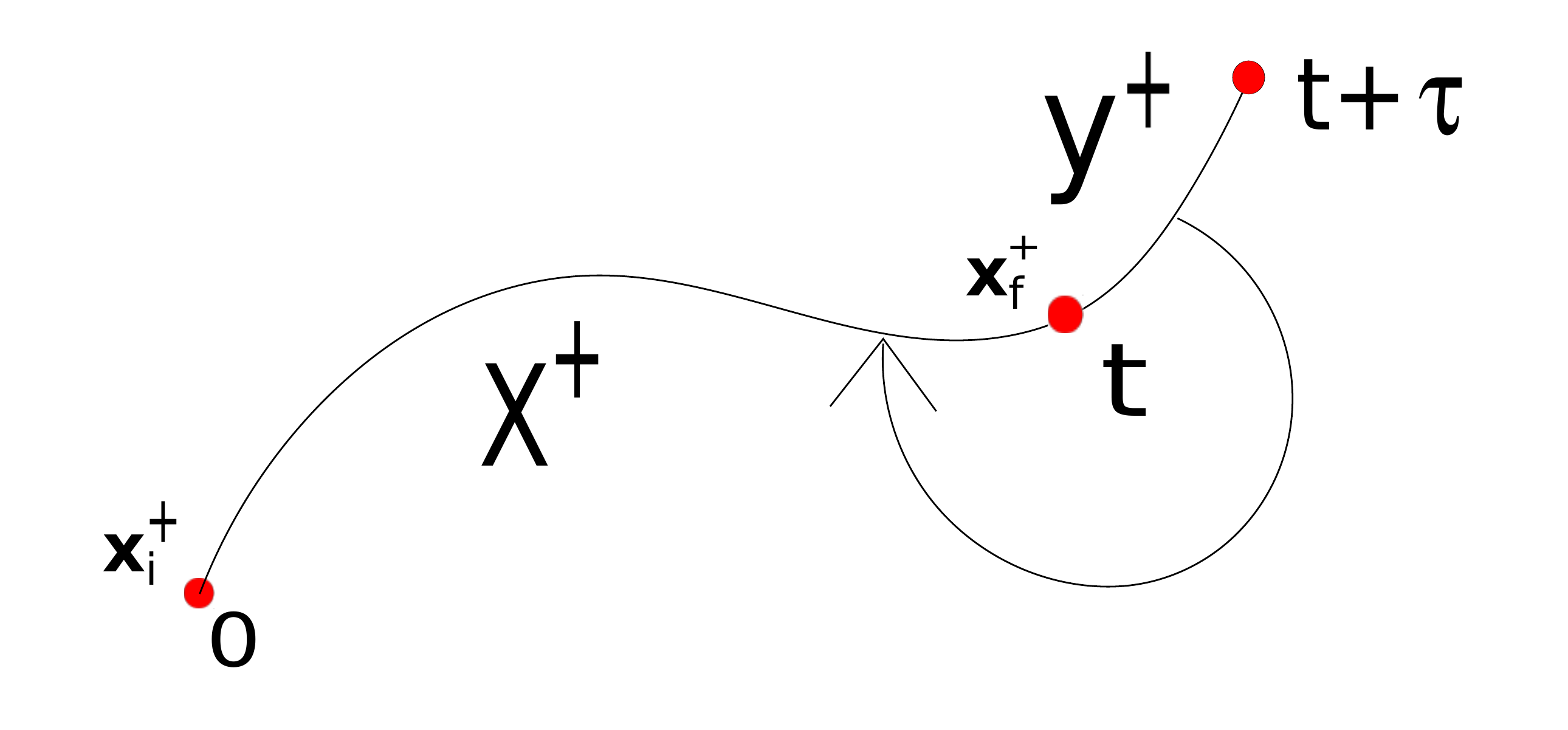}
\caption{\label{Fig1} (Color on line) Time-reversed paths ${\bf X}^\dag$ and ${\bf Y}^\dag$. When the dynamics is governed by the acausal Langevin equation (\ref{EqLtilde}), the feedback force depends on the {\it future} state of the system, as schematically represented by the arrowed line.}
\end{center}
\end{figure}

The second relation is obtained by performing the time-reversal operation normally associated with the microscopic reversibility condition~\cite{C1998}. The key point is that ${\cal Q}_t$ is no longer an odd quantity under time reversal because of the time delay. To recover this symmetry, one must also flip $\tau$ into $-\tau$ and  introduce another conjugate  dynamics defined by the {\it acausal} Langevin equation
\begin{align}
\label{EqLtilde}
m\dot  v_t=-\gamma v_t+F(x_t)+F_{bf}(x_{t+\tau})+\sqrt{2\gamma T}\xi_t\ .
\end{align}
The usual local detailed balance equation is then generalized as~\cite{MR2014,RMT2015}
\begin{align}
\label{EqELB1}
\beta  {\cal Q}_t=\ln \frac{{\cal P}[{\bf X}\vert {\bf Y}]}{\widetilde {\cal P}[{\bf X}^\dag\vert {\bf x}_i^\dag;{\bf Y}^\dag]}-\ln \frac{ {\cal J}_t}{\widetilde{\cal J}[{\bf X}]}\ ,
\end{align}
where $\widetilde {\cal P}[ {\bf X}^\dag\vert {\bf x}_i^\dag;{\bf Y}^\dag]$ is the probability of  realizing the time-reversed trajectory ${\bf X}^\dag$  with the conjugate dynamics (hereafter represented by the {\it tilde} symbol), given the initial value ${\bf x}_i^\dag \equiv {\bf x}^\dag(t=0)=(x_t,-v_t)$ and the trajectory ${\bf Y}^\dag$. Note that ${\bf Y}^\dag$ denotes the time-reversed  path in the time interval $[t,t+\tau]$, so that  its initial point  is ${\bf x}^\dag_f \equiv {\bf x}^\dag(t)=(x_0,-v_0)$, as shown schematically in Fig. \ref{Fig1}. 

Therefore, the probability weight of ${\bf X}^\dag$ must be conditioned on both the initial value ${\bf x}_i^\dag$ and the future trajectory  ${\bf Y}^\dag$. $\widetilde {\cal P}[ {\bf X}^\dag\vert  {\bf x}_i^\dag;{\bf Y}^\dag]$ is then expressed as  
\begin{align}
\label{EqtildeP}
\widetilde {\cal P}[ {\bf X}^\dag\vert  {\bf x}_i^\dag;{\bf Y}^\dag]\propto \widetilde {\cal J}[{\bf X}] e^{-\beta \widetilde {\cal S}[{\bf X}^\dag,{\bf Y}^\dag]} \ ,
\end{align}
where 
\begin{align}
\label{Eqactiontilde}
\widetilde {\cal S}[{\bf  X},{\bf Y}]=\frac{1}{4\gamma}\int_{0}^{t}dt' \:\Big[m\dot v_{t'}+\gamma v_{t'}-F(x_{t'})-F_{fb}(x_{t'+\tau})\Big]^2 \ ,
\end{align}
and $\widetilde {\cal J}[{\bf X}]$ is the  Jacobian  of the transformation $\xi(t) \rightarrow x(t)$ associated with Eq. (\ref{EqLtilde}).  As shown in I,  $\widetilde {\cal J}[{\bf X}]$ is a nontrivial functional of the path in general, but it becomes a path-independent quantity $\widetilde {\cal J}_t$ like the Jacobian ${\cal J}_t$ associated with Eq. (\ref{EqL1}) in the case of a linear dynamics. (In this work, we use the notation $\widetilde {\cal J}_t$ and ${\cal J}_t$ instead of $\widetilde {\cal J}$ and ${\cal J}$ to emphasize that these quantities depend on the duration of the trajectory.)

 Note that the two OM actions $\widetilde {\cal S}[{\bf  X},{\bf Y}]$ and $\hat {\cal S}[{\bf  X},{\bf Y}]$ are related by time inversion, namely
\begin{align}
\label{Eqactionrelation}
\widetilde {\cal S}[{\bf  X},{\bf Y}]=\hat {\cal S}[{\bf  X}^\dag,{\bf Y}^\dag] \ .
\end{align} 
 There is also an IFT associated with Eq. (\ref{EqELB1}) but it involves a more complicated path functional (see Eqs. (74)-(79) in I) that plays no role in the following. Let us just recall the corresponding second-law-like inequality for the heat flow  in the NESS~\cite{MR2014,RMT2015}:
\begin{align}
\label{Eq2law0}
 \langle \beta \dot{\cal Q}\rangle_{st}\ge -\dot {\cal S}_{\cal J}\ ,
\end{align}
where 
\begin{align}
\label{SpointJ}
\dot {\cal S}_{\cal J}\equiv \lim_{t \to \infty}\frac{1}{t}\langle\ln \frac{{\cal J}_t}{\widetilde {\cal J}[{\bf X}]}\rangle_{st}.
\end{align}
 This  bound is in general different from the trivial bound $\langle \beta  \dot {\cal Q}\rangle_{st}\ge -(\gamma/m)$ obtained from Eq. (\ref{EqQav}) (see e.g. Fig.  \ref{Fig8} in Sec. IV).

\subsection{Generating functions in the NESS}

We now focus on the steady-state regime and  drop  the suffix ``st"  in all expressions hereafter to shorten the notation.
 Our objective in this section is to express the generating functions $Z_A(\lambda,t)$ in three different ways by exploiting expressions (\ref{EqELB2}) and (\ref{EqELB1}) of the dissipated heat. 
We start from the definition (\ref{EqZ}) which we write down more explicitly as
\begin{align}
\label{EqZA0}
Z_A(\lambda,t)&=\int \int  d{\bf x}_i \:d{\bf x}_f \int d\mathbb{P}[{\bf Y}] \int_{{\bf x}_i}^{{\bf x}_f}  {\cal D}{\bf X}\: e^{-\lambda {\cal A}_t}{\cal P}[{\bf X}\vert {\bf Y}]
\end{align}
where  $\int d\mathbb{P}[{\bf Y}] ...$ is a shorthand notation for $\int d{\bf y}_i \: p({\bf y}_i ) \int_{{\bf y}_i}^{{\bf x}_i}{\cal D}{\bf Y} \: {\cal P}[{\bf Y}\vert {\bf y}_i]...$ and ${\bf y}_i\equiv (x_{-\tau},v_{-\tau})$ (hence $\int d\mathbb{P}[{\bf Y}]=p({\bf x}_i)$). Since the three observables  only differ by temporal boundary terms which are functions of $ {\bf x}_i$ and $ {\bf x}_f$,  we  single out one of them, namely $\beta {\cal W}_t$, and define the $\lambda$-dependent quantity
\begin{align}
\label{EqKlambda}
{\cal K}_{\lambda}[{\bf x}_f,t\vert {\bf Y}]&=\int_{{\bf x}_i}^{{\bf x}_f}{\cal D}{\bf X} \:  e^{-\lambda \beta {\cal W}_t} {\cal P}[{\bf X}\vert {\bf Y}]\ .
\end{align}
(The choice of $\beta {\cal W}_t$ instead of $\beta {\cal Q}_t$ or $\Sigma_t$ will be justified a posteriori in Sec. IV.B.1.) Loosely speaking, ${\cal K}_{\lambda}[{\bf x}_f,t\vert {\bf Y}]$ is a kind of biased transition probability from ${\bf x}_i$ to ${\bf x}_f$.  This allows us to re-express the three generating functions as 
\begin{align}
\label{EqZA1}
&Z_A(\lambda,t)=\int \int  d{\bf x}_i d{\bf x}_f \ f_{A,\lambda}({\bf x}_i,{\bf x}_f) \int d\mathbb{P}[{\bf Y}]\:{\cal K}_{\lambda}[{\bf x}_f,t\vert {\bf Y}]\ ,
\end{align}
where 
\begin{subequations}
\label{Eq_fA_lambda:subeqns}
\begin{align}
&f_{W,\lambda}({\bf x}_i,{\bf x}_f)=1\,, \label{Eq_fA_lambda:subeq1}\\&
f_{Q,\lambda}({\bf x}_i,{\bf x}_f)=e^{\lambda \beta \Delta {\cal U}({\bf x}_i,{\bf x}_f)} \, ,  \label{Eq_fA_lambda:subeq2}\\&
f_{\Sigma,\lambda}({\bf x}_i,{\bf x}_f)=e^{\lambda [\beta \Delta {\cal U}({\bf x}_i,{\bf x}_f)+\ln p({\bf x}_f)/p({\bf x}_i)]} \label{Eq_fA_lambda:subeq3}\ ,
\end{align}
\end{subequations}
and  we have used the first law (\ref{Eq1law}) to define $f_{Q,\lambda}$ and $f_{\Sigma,\lambda}$. 

We now use Eq. (\ref{EqELB2})  to replace  the  path probability $ {\cal P}[{\bf X}\vert {\bf Y}]$ by $
\hat{\cal P}[{\bf X}\vert {\bf Y}]$ in Eq. (\ref{EqZA0}). We then define
\begin{align}
\label{EqKhat}
\hat {\cal K}_{\lambda}[{\bf x}_f,t\vert {\bf Y}]&=e^{\frac{\gamma }{m}t}\int_{{\bf x}_i}^{{\bf x}_f}{\cal D}{\bf X} \:  e^{(1-\lambda) \beta{\cal W}_t}\hat {\cal P}[{\bf X}\vert {\bf Y}]\ ,
\end{align}
which leads to 
\begin{align}
\label{EqZA3}
Z_A(\lambda,t)&= \int \int  d{\bf x}_i  d{\bf x}_f \hat f_{A,\lambda}({\bf x}_i,{\bf x}_f)\int d\mathbb{P}[{\bf Y}]\:\hat{\cal K}_{\lambda}[{\bf x}_f,t\vert {\bf Y}]\ ,
\end{align}
with
\begin{subequations}
\label{Eq_fA_lambda3:subeqns}
\begin{align}
&\hat f_{W,\lambda}({\bf x}_i,{\bf x}_f)=e^{-\beta \Delta {\cal U}({\bf x}_i,{\bf x}_f)}\,, \label{Eq_fA_lambda3:subeqns1}\\&
\hat f_{Q,\lambda}({\bf x}_i,{\bf x}_f)=e^{(\lambda-1)\beta \Delta {\cal U}({\bf x}_i,{\bf x}_f)} \, ,  \label{Eq_fA_lambda3:subeqns2}\\&
\hat f_{\Sigma,\lambda}({\bf x}_i,{\bf x}_f)= e^{(\lambda-1)\beta \Delta {\cal U}({\bf x}_i,{\bf x}_f)+\lambda \ln \frac{p({\bf x}_f)}{p({\bf x}_i)}} \ .\label{Eq_fA_lambda3:subeqns3}
\end{align}
\end{subequations}

Likewise, we  can use  Eq. (\ref{EqELB1}) to replace  ${\cal P}[ {\bf X}\vert {\bf Y}]$ by $\widetilde {\cal P}[ {\bf X}^\dag\vert {\bf x}_i^\dag;{\bf Y}^\dag]$. In this case, it is  convenient to change the path integral over ${\bf X}$ in Eq. (\ref{EqZA0}) into an integral over  ${\bf X}^\dag$, and  define
\begin{align}
\label{EqKtilde}
\widetilde{\cal K}_{\lambda}({\bf x}_f^\dag,t\vert {\bf x}_i^\dag; {\bf Y}^\dag)= \int_{{\bf x}_i^\dag}^{{\bf x}_f^\dag} {\cal D}{\bf X}^\dag\: \frac{{\cal J}_t}{\widetilde {\cal J}[{\bf X}]}e^{(1-\lambda)\beta {\cal W}_t} \widetilde {\cal P}[{\bf X}^\dag\vert {\bf x}_i^\dag,{\bf Y}^\dag]\ ,
\end{align}
which leads to
\begin{align}
\label{EqZA4}
Z_A(\lambda,t)&= \int \int  d{\bf x}_i^\dag  d{\bf x}_f^\dag \:\tilde f_{A,\lambda}({\bf x}_i,{\bf x}_f) 
\nonumber\\
&\times \int d\mathbb{P}[{\bf Y}]\:\widetilde{\cal K}_{\lambda}({\bf x}_f^\dag,t\vert {\bf x}_i^\dag; {\bf Y}^\dag)\ ,
\end{align}
with 
\begin{align}
\widetilde f_{A,\lambda}({\bf x}_i,{\bf x}_f)=\hat f_{A,\lambda}({\bf x}_i,{\bf x}_f)\,.
\end{align}

At first glance, it might seem that we have gained nothing by replacing Eq. (\ref{EqZA1}) by two other expressions of the generating functions that are even more complicated. This is  true for a generic value of $\lambda$. But the interesting feature of  Eqs. (\ref{EqZA3}) and (\ref{EqZA4})  is the special role played by $\lambda=1$. This makes these two equations well suited to infer the asymptotic behavior of the quantities $Z_A(1,t)=\langle e^{-{\cal A}_t}\rangle$, thus revealing the occurrence of large statistical fluctuations originating from temporal boundary terms. However, this requires to first determine whether or not the conjugate Langevin equations (\ref{EqL1minus})  and (\ref{EqLtilde}) admit a stationary solution. Although this analysis can be done in a rather general framework, it is quite delicate in the case of the acausal dynamics (\ref{EqLtilde}) and it is more illuminating to focus on a specific case, namely the linear model studied in the next section.  We shall thus return to this issue later on. A detailed discussion is presented in Appendix A.

\section{Time-delayed Langevin harmonic oscillator}

To be concrete, we now consider the  time-delayed linear Langevin equation
\begin{align}
\label{EqL}
m\dot v_t=-\gamma v_t-kx_t+k'x_{t-\tau}+\sqrt{2\gamma T}\xi_t
\end{align}
which is conveniently rewritten in a dimensionless form as
 \begin{align}
\label{EqLlinred}
\dot v_t=-\frac{1}{Q_0} v_t -x_t+\frac{g}{Q_0}x_{t-\tau}+\xi_t
\end{align}
by taking the inverse angular resonance frequency $\omega_0^{-1}= \sqrt{m/k}$ as the unit of time and $x_c=k^{-1}(2\gamma T)^{1/2}$ as the unit of position~\cite{NF2011}. In this equation, $Q_0=\omega_0 \tau_0$ denotes the intrinsic quality factor of the oscillator ($\tau_0=m/\gamma$ is the viscous relaxation time) and $g=k'/(\gamma \omega_0)=(k'/k)Q_0$ represents the gain of the feedback loop. The  dynamics of the system is thus fully characterized by the three independent dimensionless parameters $Q_0$, $g$ and $\tau$. The gain $g$ is usually the control variable in feedback-cooling experimental setups (see e.g. ~\cite{PZ2012}).

In these reduced units, the fluctuating work and heat  (normalized by $k_BT$) are given by
\begin{align}
\label{EqWXYred}
\beta {\cal W}_t=\frac{2g}{Q_0^2}\int_{0}^{t} dt' \: x_{t'-\tau}\circ v_{t'}\ ,
\end{align}
and
\begin{align}
\label{EqQXYred}
\beta {\cal Q}_t=\beta {\cal W}_t-\frac{1}{Q_0}(x_f^2-x_i^2+v_f^2-v_i^2)\ .
\end{align}
These are quadratic functionals of the noise and therefore the corresponding probabilities are not Gaussian. To obtain the expression of the EP functional $\Sigma_t$ defined by Eq. (\ref{EqSigmaXY}), we use the expression of the stationary pdf derived in I,
\begin{align}
\label{Eqpdf1}
p({\bf x})&=\frac{1}{2\pi[\langle x^2\rangle\langle v^2\rangle]^{1/2}}e^{-\frac{x^2}{2\langle x^2\rangle}-\frac{v^2}{2\langle v^2\rangle}}\ ,
\end{align}
where the mean square position and velocity are expressed in terms of the configurational  and kinetic temperatures $T_x$ and $T_v$: $\langle x^2\rangle=(Q_0/2) T_x/T$ and $\langle v^2\rangle=(Q_0/2) T_v/T$. These two effective temperatures are given by Eqs. (113) and (114) in I, respectively. We recall that $T_x$ is the temperature commonly measured in experiments involving nanomechanical devices~\cite{PCBH2000,Mont2012,PDMR2007} whereas $T_v$ determines the heat flow (and thus the extracted work) in the stationary state, according to
\begin{align}
\label{Eqavheat}
\beta \dot {\cal Q}=\frac{1}{Q_0}(\frac{T_v}{T}-1)\ ,
\end{align}
i.e., $\beta \dot {\cal Q}=(\gamma/m)(T_v/T-1)$ in original units (see Eq. (\ref{EqQav})). Since $T_x\ne T_v$ in general, the system does not obey the standard equipartition theorem (cf. the discussion in I).
 Inserting Eq. (\ref{Eqpdf1}) into Eq. (\ref{EqSigmaXY}) then yields
\begin{align}
\label{EqSXYred}
\beta \Sigma_t&=\beta {\cal W}_t+\frac{1}{Q_0} \left[\frac{T-T_x}{T_x}(x_t^2-x_0^2)+\frac{T-T_v}{T_v}(v_t^2-v_0^2)\right]\ .
\end{align}
 
Since we are dealing with a linear dynamics with Gaussian noise, all stationary path probabilities are Gaussian distributions~\cite{note5} and the calculation of the generating functions $Z_A(\lambda,t)$ from Eqs. (\ref{EqZA1}) and (\ref{Eq_fA_lambda:subeqns}) amounts to computing Gaussian path integrals. For $t$ finite, however, this calculation cannot be carried out analytically for essentially two reasons. The first is that the  Euler-Lagrange equation for the optimal trajectory is a forward-backward delay differential equation that has no closed-form solution in general (it can only be solved by a perturbative expansion in powers of $g$). The second reason is that the explicit expression of ${\cal P}[{\bf Y}]$ is unknown~\cite{note120}, so that the average over initial conditions cannot be performed.

Things become simpler in the long-time limit as one can use the Fourier transform to obtain an analytical expression of the SCGF  $\mu(\lambda)$, see Eq. (\ref{Eqmulambda}) below (but, as already stressed, the value at $\lambda=1$ requires  special care).  The LDFs are then obtained via the Legendre transform in Eq. (\ref{EqLT}). However, since the behavior of the prefactors $g_A(\lambda)$ is unknown and singularities may occur, additional assumptions are needed. Useful insight on this issue is gained by inspecting the small-$\tau$ limit of the Langevin equation, which corresponds to the Markovian model originally considered in \cite{KQ2004,KQ2007}. The feedback then generates an additional viscous damping  and a complete  analytical description of the fluctuations is possible, as detailed in  Appendix B.  (A first, but incomplete analysis was performed by two of us in \cite{MR2012}.) This study, together with the additional pieces of information gathered from the direct numerical simulation of Eq. (\ref{EqL}), will eventually allow us to propose a global scenario.

In order to give the reader a foretaste of the puzzle that must be resolved, we first present some data obtained from numerical simulations of the Langevin equation ({\ref{EqLlinred}). The theoretical interpretation is postponed to Sec. IV B.

\subsection {Numerical study}

\begin{figure}[hbt]
\begin{center}
\includegraphics[trim={0 1.25cm 0 1.75cm},clip,width=8cm]{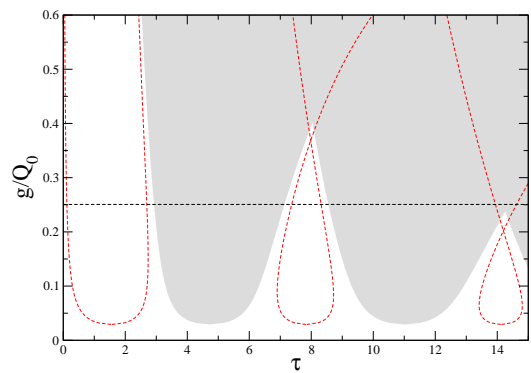}
 \caption{\label{Fig2} (Color on line) Stability diagram  of the feedback-controlled oscillator for $Q_0=34.2$ (the time unit is the inverse angular resonance frequency $\omega_0^{-1}$). The oscillator is unstable inside the shaded regions. The acausal response function $\widetilde \chi(s)$ has all its poles located in the r.h.s. of the complex $s$-plane inside the regions delimited by the dashed red lines and two poles in the l.h.s. outside these regions.}
\end{center}
\end{figure}

Although we have studied the model  for various values of the dimensionless parameters $Q_0$ and $g$,  we  here only present numerical results obtained for $Q_0=34.2$ and $g/Q_0=0.25$. We have chosen this set of  parameters for several reasons. In the first place, the value of $Q_0$ corresponds to an actual experimental system: the AFM micro-cantilever used in the experiments of Ref.~\cite{GBPC2010}, which is characterized by a resonance period $2\pi/\omega_0=116 \: \mu s$ and a viscous relaxation time $\tau_0=632\: \mu s$. In the second place,  the feedback-controlled oscillator has an interesting dynamical behavior for $g/Q_0=0.25$, as shown in Fig. \ref{Fig2} (see also Fig. 11 in I). Specifically, a  stationary state can be reached in  two  stability lobes $0< \tau<\tau_1$ and $\tau_2< \tau<\tau_3$, with $\tau_1\approx 2.93$, $\tau_2\approx 7.13$, $\tau_3\approx 8.55$. For intermediate values of $\tau$ or  $\tau >\tau_3$, there is no stationary state.  In the third place, the probability distributions  have a nontrivial behavior as a function of $\tau$ which vividly illustrates the role of rare events due to boundary temporal terms.
\begin{figure}[hbt]
\begin{center}
\includegraphics[trim={0 1.25cm 0 1.75cm},clip,width=8cm]{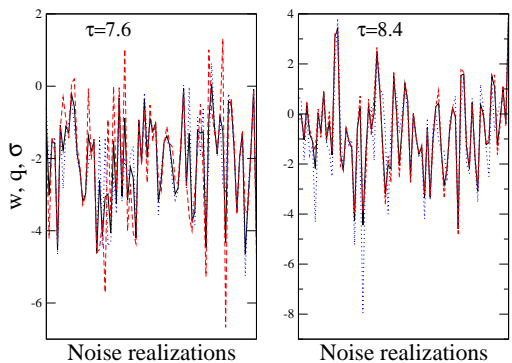}
 \caption{\label{Fig3} (Color on line) Stochastic fluctuations of  $w=\beta{\cal W}_t/t$ (solid black line), $q=\beta {\cal Q}_t/t$ (dotted blue line), and $\sigma=\Sigma_t/t$ (dashed red line)  for $Q_0=34.2$ $g/Q_0=0.25$, $\tau= 7.6$ (left panel) and $\tau=8.4$ (right panel). The figure shows the results obtained for an observation time $t=100$ and $75$ independent noise realizations. Lines are only a guide for the eyes.}
\end{center}
\end{figure}

To begin with, we show in Fig. \ref{Fig3} an example of the sample-to-sample fluctuations of $\beta {\cal W}_t$, $\beta {\cal Q}_t$, and $\Sigma_t$ for two values of $\tau$ chosen in the second stability lobe $\tau_2< \tau<\tau_3$ (the behavior is qualitatively similar in the first lobe). The  observation time is $t=100$ and the Langevin equation (\ref{EqLlinred}) was solved by using Heun's method~\cite{M2002} with a time-step $\Delta t=5. 10^{-4}$. 
As expected, the fluctuations of the three observables are strongly correlated. But, remarkably, the contribution of the temporal boundary terms is still non negligible despite the long observation time. In particular, they contribute differently to the observables depending on the value of $\tau$: for $\tau=7.6$ (resp. $\tau=8.4$) it is $\Sigma_t$ (resp. ${\cal Q}_t$) that exhibits the largest fluctuations.  Note that the delay is significantly smaller than the viscous relaxation time $\tau_0=Q_0/\omega_0=34.2$ in both cases, and that the system operates in the  cooling regime: $T_x/T\approx 0.42$, $T_v/T\approx 0.36$, $\langle \beta \dot {\cal Q}\rangle \approx -0.019$ for $\tau=7.6$, and $T_x/T\approx 0.72$, $T_v/T\approx 0.84$, $\langle \beta \dot {\cal Q}\rangle \approx -0.005$ for $\tau=8.4$. 

\begin{figure}[hbt]
\begin{center}new
\includegraphics[trim={0 1.25cm 0 1.75cm},clip,width=8cm]{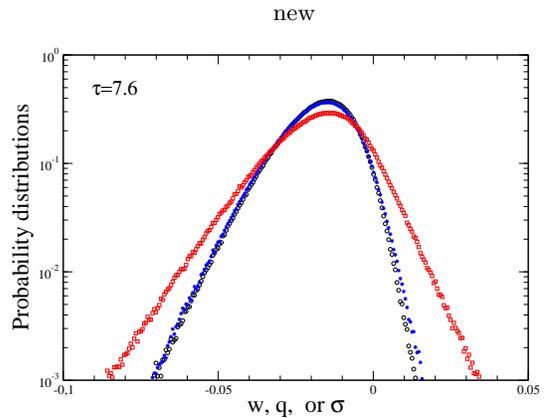}
 \caption{\label{Fig4} (Color on line)  Probability distribution functions $P(\beta {\cal W}_t)$ (black circles), $P(\beta {\cal Q}_t)$ (blue stars), and $P(\Sigma_t)$ (red squares)  plotted against the scaled variable $a= {\cal A}_t/t$ ($a=w,q$ or $\sigma$) for $Q_0=34.2$, $g/Q_0=0.25$ and $\tau=7.6$. The observation time is $t=100$. Symbols represent numerical data obtained by solving the Langevin equation (\ref{EqLlinred}) for $2.10^6$ realizations of the noise.}
\end{center}
\end{figure}
\begin{figure}[hbt]
\begin{center}
\includegraphics[trim={0 1.25cm 0 1.75cm},clip,width=8cm]{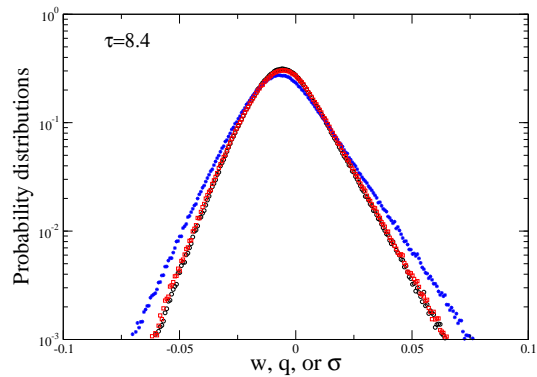}
 \caption{\label{Fig5} (Color on line) Same as Fig. \ref{Fig3} for $\tau=8.4$.}
\end{center}
\end{figure}
To get a more quantitative picture,  the corresponding stationary pdfs are shown  in  Figs. \ref{Fig4} and \ref{Fig5}. These plots clearly confirm the main feature suggested by Fig. \ref{Fig3}:  $P(\Sigma_t=\sigma t)$  for $\tau=7.6$ and $P(\beta{\cal Q}_t=qt)$ for $\tau=8.4$ differ markedly  from $P(\beta {\cal W}_t=wt)$. Of course, these results should be interpreted with care since it is notoriously difficult to sample rare fluctuations.  However, we expect that the picture emerging from Figs.  \ref{Fig4}  and \ref{Fig5}  would not change qualitatively at larger times. Moreover, it is  consistent with the exact analytical analysis performed in Appendix B in the small-$\tau$ limit and in the associated Markovian model. This will be rationalized in the next subsection.

The corresponding  estimates of the SCGFs $\mu_A(\lambda)$ are plotted in Fig. \ref{Fig6}. One  
 noticeable feature is the distinct behavior of $\mu_{\Sigma}(\lambda)$ for $\tau=7.6$ and of $\mu_Q(\lambda)$  for  $\tau=8.4$ in the vicinity of $\lambda=1$. However, it is also manifest that finite-time and/or finite-sample-size effects are significant. In particular, $\mu_{\Sigma}(\lambda)$  for $\tau=7.6$ widely differs from the two other SCGFs for $\lambda\lessapprox -1$ and varies linearly with $\lambda$ for $\lambda\lessapprox -1.5$, which is presumably a numerical artifact, as discussed in a more general context in \cite{RAT2015}. 
\begin{figure}[hbt]
\begin{center}
\includegraphics[trim={0 1.25cm 0 1.75cm},clip,width=8cm]{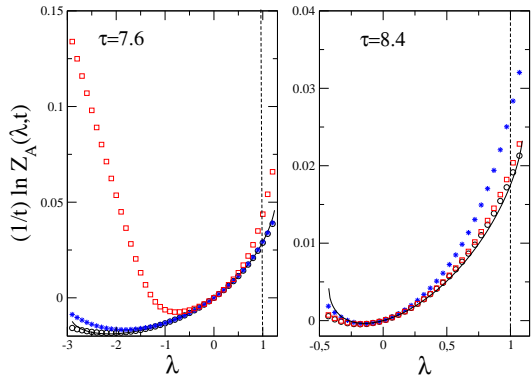}
 \caption{\label{Fig6} (Color on line) Numerical estimates of $\mu_A(\lambda)$ using $t=100$ and $2.10^6$ realizations of the noise: $\mu_W$ (black circles), $\mu_Q$ (blue stars), and $\mu_{\Sigma}$ (red squares). The solid black line represents  the theoretical SCGF $\mu(\lambda)$ given by Eq. (\ref{Eqmulambda}) in the interval $(\lambda_{min},\lambda_{max})$ in which this quantity is real.}
\end{center}
\end{figure}

Finally, we focus on the special value $\lambda=1$ and show in Fig. \ref{Fig7} the numerical estimates of  $\mu_A(1)=\lim_{t\rightarrow \infty} (1/t)\ln\langle e^{- {\cal A}_t}\rangle$ in the whole  stability range $7.13<\tau<8.55$. For information, we also show the average extracted work rate $\langle \beta \dot {\cal W}_{ext}\rangle =-\langle \beta \dot {\cal Q}\rangle $. We first observe that $\mu_Q(1)\approx  1/Q_0$ ($\gamma/m$ in original units) independently of the value of $\tau$. This is indeed what the  IFT (\ref{EqIFTQ}) tells us. In contrast, both $\mu_W(1)$ and $\mu_{\Sigma}(1)$ display a nontrivial behavior with $\tau$: $\mu_W(1)$ is equal to $1/Q_0$ only the sub-interval in $7.37\lessapprox \tau\lessapprox 8.32$, whereas $\mu_{\Sigma}(1)$ varies nonmonotonically with a maximum around $\tau\approx 7.9$. This clearly calls for a theoretical explanation. 
\begin{figure}[hbt]
\begin{center}
\includegraphics[trim={0 1.25cm 0 1.75cm},clip,width=8cm]{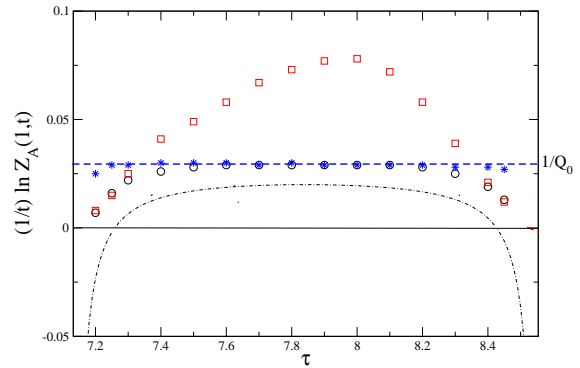}
 \caption{\label{Fig7} (Color on line) Numerical estimates of $ \mu_A(1)=\lim_{t\rightarrow \infty} (1/t)\ln\langle e^{- {\cal A}_t}\rangle$ as a function of $\tau$ in the second stability lobe: $\mu_W(1)$ (black circles), $\mu_Q(1)$ (blue stars), and $\mu_{\Sigma}(1)$ (red squares). 
 The dashed-dotted black line is the average extracted work rate $\langle \beta \dot {\cal W}_{ext}\rangle =-\langle \beta \dot {\cal Q}\rangle $ which is positive for $7.26\le \tau \le 8.43$.}
\end{center}
\end{figure}

\subsection{Theoretical analysis}

We now present a theoretical scenario that explains why  the three observables, which only differ by temporal boundary terms, have different fluctuations in the long-time limit and why this behavior depends on the delay. 

\subsubsection{Calculation of the (boundary-independent) SCGF}

We first calculate the SCGF $\mu(\lambda)$ defined by the asymptotic formula (\ref{EqZ0}), assuming that boundary terms depending on ${\bf x}_i,{\bf x}_f$ or ${\bf Y}$ play no role in the long-time limit. (We recall that the trajectory ${\bf Y}$ is of duration $\tau$.)  However, one should keep in mind that the actual value of $\mu_A(1)$  may differ from $\mu(1)$. 

In order to compute $\mu(\lambda)$, we impose periodic boundary conditions on the trajectory ${\bf X}$ and expand $x_t$ in discrete Fourier series (see e.g. \cite{ZBCK2005,KSD2011,SD2011,FI2012,RH2016} for similar calculations),
\begin{align} 
x(t)&=\sum_{n=-\infty}^{\infty} x_ne^{-i\omega_n t}\ ,
\end{align}
with  inverse transform
\begin{align} 
x_n&=\frac{1}{t} \int_0^t ds \:x(s) e^{i\omega_n s}
\end{align}
where $\omega_n=2\pi n/t$ and $x_n\equiv x(\omega_n)$.  
 In the limit $t\rightarrow \infty$,  the standard Fourier transform is recovered.

After inserting into Eq. (\ref{EqWXYred}) and neglecting the contribution coming from ${\bf Y}$, we obtain
 \begin{align}
\label{EqwFourier}
\frac{1}{t} \beta {\cal W}_t&\sim\frac{2g}{Q_0^2}\sum_{n=-\infty}^{\infty}(i\omega_n)  x_n  x_{-n}e^{i\omega_n\tau}\nonumber\\
&\sim -\frac{4g}{Q_0^2}\sum_{n=1}^{\infty}\omega_n  x_n  x_{-n}\sin(\omega_n\tau)\ .
\end{align}
(Of course, this expression is also valid for $\beta {\cal Q}_t$ or $\Sigma_t$ since the contribution of the boundary terms are neglected in this calculation.) We then use the linearity of the Langevin equation to replace $x_n$ by the frequency component $\xi_n$ of the noise
 \begin{align}
 x_n=\chi(\omega_n) \xi_n\ ,
\end{align}
where
 \begin{align}
\label{Eqchi}
\chi(\omega)=[-\omega^2-\frac{i\omega}{Q_0}+1-\frac{g}{Q_0}e^{i\omega\tau}]^{-1}
\end{align}
is the Fourier transform of the response function of the time-delayed oscillator. Hence,
 \begin{align}
\langle e^{-\lambda \beta{\cal W}_t}\rangle &\sim \prod_{n=1}^{\infty}\int d\xi_n\: P(\xi_n)e^{ \frac{4\lambda g t}{Q_0^2}\xi_n  \xi_{-n}\omega_n \sin(\omega_n\tau) \vert\chi(\omega_n)\vert^2}
\end{align}
with
\begin{align}
{\cal P}[\xi_n]=\frac{t}{\pi}e^{-t \xi_n  \xi_{-n}} \ .
\end{align}
The Gaussian integration over $\xi_n$  gives
 \begin{align}
\langle e^{-\lambda \beta{\cal W}_t}\rangle &\sim \prod_{n=1}^{\infty} [1-\frac{4\lambda g}{Q_0^2}\omega_n \sin(\omega_n\tau) \vert\chi(\omega_n)\vert^2]^{-1}\ ,
\end{align}
and we finally obtain
\begin{align}
\label{Eqmulambda}
\mu(\lambda)&=-\lim_{t\rightarrow \infty}\frac{1}{t}\sum_{n=1}^{\infty}\ln [1-\frac{4\lambda g}{Q_0^2}\omega_n \sin(\omega_n\tau) \vert \chi(\omega_n)\vert ^2 ]\nonumber\\
&=- \frac{1}{2\pi}\int_{0}^{\infty}d\omega\: \ln [1-\frac{4\lambda g}{Q_0^2}\omega \sin(\omega \tau) \vert \chi(\omega)\vert ^2 ] \ ,
\end{align}
where the summation over $n$ has been replaced by an integral over $\omega$ as $t \to \infty$. For a generic value of $\lambda$, the  integral  must be computed numerically, and the result is real as long as the argument of the logarithm stays positive for all values of $\omega$. Accordingly, $\mu(\lambda)$ is defined in an open domain $(\lambda_{min},\lambda_{max})$, with  $\lambda_{min}$ and $\lambda_{max}$ determined  by the minimum and maximum values of the function $f(\omega)=(4g/Q_0^2)\omega \sin(\omega \tau) \vert\chi(\omega)\vert ^{2}$. The derivative $\mu'(\lambda)$ diverges at the boundaries, so that the corresponding Legendre transform is  asymptotically linear~\cite{T2009}.

As regards fluctuation relations, we readily notice  from  Eq. (\ref{Eqmulambda}) that  $\mu(1-\lambda)\ne \mu(\lambda)$, which implies that the observables do not satisfy a conventional stationary-state fluctuation theorem (SSFT) of the Gallavotti-Cohen type~\cite{GC1995,K1998,LS1999}: $\lim_{t\rightarrow \infty}(1/t)\ln [P({\cal A}_t=at)/P({\cal A}_t=-at)]=a$. On the other hand, alternative SSFTs can be obtained by  changing $\gamma$ into $-\gamma$ or $\tau$ into $-\tau$. We will say more about this in subsection B.4.

How does Eq. (\ref{Eqmulambda})  compare with the numerical estimates of the SCGFs $\mu_A(\lambda)$ shown in Fig. \ref{Fig6} ? We  see that the agreement is very good for $\mu_W(\lambda)$, although there are still small discrepancies, in particular for $\tau=8.4$ and  the most negative values of $\lambda$. These  small deviations will be used to infer the numerical value of the prefactor $g_W(\lambda)$ and build a better approximation of the  pdf $P(\beta {\cal W}_t=wt)$ (see Eq. (\ref{EqPWfull}) below). Much more significant are the differences with $\mu_{\Sigma}(\lambda)$  for $\tau=7.6$ and with $\mu_Q(\lambda)$ for  $\tau=8.4$ in the vicinity of $\lambda=1$ (leaving aside the spurious linear behavior of  $\mu_{\Sigma}(\lambda)$  for $\tau=7.6$ and $\lambda\lesssim -1.5$).

Let us investigate this issue in more detail by computing $\mu(1)$. To this aim, we  first rewrite Eq. (\ref{Eqmulambda}) as
\begin{align}
\label{Eqmuw1}
\mu(\lambda)=  \frac{1}{2\pi}\int_{0}^{\infty} d\omega\: \ln \frac{H_{\lambda}(\omega)}{H_0(\omega)}\ ,
\end{align}
where 
\begin{align}
\label{Hlambda}
H_{\lambda}(\omega)^{-1}&\equiv \vert\chi(\omega)\vert ^{-2}-\frac{4\lambda g}{Q_0^2}\omega \sin(\omega \tau)\nonumber\\
&= [-\omega^2+1-\frac{g}{Q_0}\cos (\omega \tau)]^2\nonumber\\
&+\frac{1}{Q_0^2}[\omega^2+2g(1-2\lambda)\omega \sin(\omega \tau)+g^2\sin^2 (\omega \tau)] 
\end{align}
and $H_0(\omega)=\vert \chi(\omega)\vert ^2$. This immediately shows that
\begin{align}
\label{Hlambda1}
H_{1}(\omega)^{-1}&= [-\omega^2+1-\frac{g}{Q_0}\cos (\omega \tau)]^2+\frac{1}{Q_0^2}[\omega-g\sin(\omega \tau)]^2 \nonumber\\
&\equiv \vert \widetilde \chi(\omega)\vert^{-2}\ ,
\end{align}
where
\begin{align}
\widetilde \chi(\omega)\equiv \chi(\omega)\vert_{\tau \rightarrow -\tau}=[-\omega^2-\frac{i\omega}{Q_0}+1-\frac{g}{Q_0}e^{-i\omega\tau}]^{-1} 
\end{align}
is the response function of the acausal Langevin equation in the frequency domain. This allows us to express $\mu(1)$ as\begin{align}
\label{Eqmuw2}
\mu(1)&= \frac{1}{2\pi}\int_{0}^{+\infty}d\omega \: \ln \frac{\vert \widetilde \chi(\omega)\vert^2}{\vert \chi(\omega)\vert^2}= \frac{1}{2\pi}\int_{-\infty}^{+\infty}d\omega \: \ln \frac{\widetilde \chi(\omega)}{\chi(\omega)}\ ,
\end{align}
where we have used the fact that the imaginary parts of $\widetilde \chi(\omega)$ and $\chi(\omega)$ are odd functions of $\omega$ to eliminate the modulus~\cite{note35}. We can then compute the integral over $\omega$ by using Cauchy's residue theorem, which requires to locate the poles  of $\widetilde \chi(\omega)$ in the complex frequency plane (they are not restricted to be in the lower half plane, in contrast with the poles of  the causal response function $\chi(\omega)$). Fortunately, this nontrivial task has already been accomplished in I in order to calculate the quantity $\dot {\cal S}_{\cal J}\equiv \lim_{t\rightarrow \infty} (1/t)\ln {\cal J}_t/\widetilde {\cal J}_t$ 
involved in the second-law-like  inequality (\ref{Eq2law0}) obtained from time reversal (we recall that the Jacobian $\widetilde {\cal J}[{\bf X}]$ becomes a path-independent quantity $\widetilde{\cal J}_t$ when the dynamics is linear~\cite{RMT2015}). Specifically, it was shown in I [Eq. (155)] that  
\begin{align}
\label{EqSdotJ}
\dot {\cal S}_{\cal J}=\frac{1}{2\pi i}\int_{c-i\infty}^{c+i\infty}ds \: \ln \frac{\widetilde \chi(s)}{\chi_{g=0}(s)} \ ,
\end{align}
where $s=\sigma -i\omega$ is the Laplace complex variable. (From now on, we will mostly work with the Laplace variable in order to directly use the results obtained in I, but for simplicity we will keep the same notation for the response functions.)  We thus re-express Eq. (\ref{Eqmuw2}) as
\begin{align}
\label{Eqmuw3}
\mu(1)&= \frac{1}{2\pi i}\int_{0-i\infty}^{0+i\infty}ds \: \ln \frac{\widetilde \chi(s)}{\chi(s)}\ ,
\end{align}
where
\begin{align}
\chi(s)=[s^2+\frac{s}{Q_0}+1-\frac{g}{Q_0}e^{-s\tau}]^{-1} \ .
\end{align}
and $\widetilde \chi(s)\equiv \chi(s)\vert_{\tau \rightarrow -\tau}$. 
Comparing  Eq.  (\ref{Eqmuw3}) to Eq. (\ref {EqSdotJ}), one may notice  two differences: firstly, one has  $\chi(s)$ in the  denominator of the logarithm instead of $\chi_{g=0}(s)$, and secondly, the integration  is performed along  the imaginary axis $Re(s)=0$ in the complex $s$-plane (since the frequency $\omega$ is  real). On the other hand, as was painstakingly discussed in I, the Bromwich contour in Eq. (\ref{EqSdotJ}) (i.e., the value of $c$) crucially depends on the location of the poles of $\widetilde \chi(s)$. The first difference turns out to be irrelevant because all the poles of $\chi(s)$ are  located in the left-hand-side (l.h.s.) of the complex $s$-plane. Hence~\cite{note40},
\begin{align}
&\frac{1}{2\pi i}\int_{0-i\infty}^{0+i\infty}ds \:\ln \frac{\chi(s)}{\chi_{g=0}(s)}=0 \ .
\end{align}
On the other hand, the fact that $c=0$ in Eq. (\ref{Eqmuw3}) is relevant in two circumstances: 

1) When all the poles of $\widetilde \chi(s)$ lie on the right-hand-side (r.h.s.) of the complex $s$-plane. Then, by using an integration contour similar to the one in Fig. 4 of I (with a large semi-circle on the l.h.s.),  the only singularities inside the contour are the two poles of  $\chi_{g=0}(s)$, $s_0^{\pm}=1/(2Q_0)[-1\pm i\sqrt{4Q_0^2-1}]$. Cauchy's residue theorem then gives the simple result $\mu(1)=1/Q_0$. This differs from $\dot S_{\cal J}$ because this latter quantity is obtained by also including two poles of $\widetilde \chi(s)$ inside the contour in order to avoid the branch cuts of the logarithm. Indeed, as shown in I, there must be two, and only two, poles of $\widetilde \chi(s)$ on the left side of the integration line $Re(s)=c$.  

2) When $\widetilde \chi(s)$ has more than two poles on the l.h.s. In this case, all these poles contribute to Eq. (\ref{Eqmuw3}) whereas only the two poles  with the smallest real part contribute to $\dot S_{\cal J}$.

To sum up,  three different cases may occur:
\begin{enumerate}[(a)]
 \item $\mu(1)=1/Q_0\ne \dot S_{\cal J}$ when all the poles of $\widetilde \chi(s)$ lie on the r.h.s of the complex $s$-plane,
 \item  $\mu(1)=\dot S_{\cal J}\ne 1/Q_0$ when  only two poles lie on the l.h.s of the complex $s$-plane,
 \item  $\mu(1)\ne \dot S_{\cal J}\ne 1/Q_0$ when more than two poles lie on the l.h.s of the complex $s$-plane~\cite{note45}.
\end{enumerate}

This calculation of $\mu(1)$, combined with the analysis performed in Appendix A, allows us to elucidate the intriguing dependence of $\mu_W(1)$ and $\mu_{\Sigma}(1)$ on $\tau$ exhibited in Fig. \ref{Fig7}, and more generally the behavior of the SCGFs in the vicinity of $\lambda=1$ observed in Fig. \ref{Fig6}. What is done in Appendix A is first to relate the behavior of the conjugate  $\gamma \to -\gamma$ ``hat"  and $\tau \to -\tau$ ``tilde" dynamics to the pole structure of $\widetilde \chi(s)$. Then, in a second time, Eqs. (\ref{EqZA3}) and (\ref{EqZA4}) derived in Sec. III.B are  used to deduce the values of $\mu_W(1)$ and $\mu_{\Sigma}(1)$. Specifically, it is shown that a stationary state exists with the hat dynamics when all the poles of $\widetilde \chi(s)$ are on the r.h.s of the complex $s$-plane [case (a)  above] and with the tilde dynamics when two (and only two) poles are on the l.h.s  [case (b)]. (When there are more than two poles on the l.h.s. [case (c)], a stationary state never exists.) With our present choice for the quality factor $Q_0$ and the feedback gain $g$ ($Q_0=34.2$ and $g/Q_0=0.25$), we find that case (a) is realized for $\tau_{p,1}<\tau<\tau_{p,2}$, with $\tau_{p,1}\approx 7.37,\tau_{p,2}\approx 8.32$, and case (b) is realized for $\tau <\tau_{p,1}$ and $\tau >\tau_{p,2}$  (for other values of $g$,  the boundary between cases (a) and (b) is indicated by the dashed red lines in  Fig. \ref{Fig2}).  

The analysis in Appendix A then tells us that 
\begin{align}
\mu_W(1)= \mu(1)
\end{align}
in {\it both} cases, i.e.,  in the whole stability lobe, as illustrated by the solid black line in Fig. \ref{Fig8}, whereas
\begin{align}
\mu_{\Sigma}(1)&=\dot S_{\cal J}=\mu(1) \ \ \mbox{for} \ \tau <\tau_{p,1} \ \mbox{and}\  \tau >\tau_{p,2}\ .
\end{align}
\begin{figure}
\begin{center}
\includegraphics[trim={0 1.25cm 0 1.75cm},clip,width=8cm]{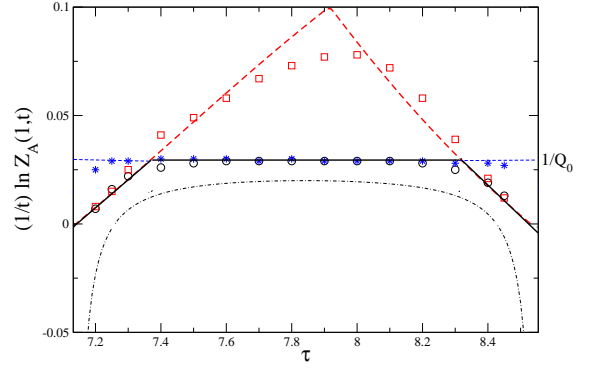}
 \caption{\label{Fig8} (Color on line) Comparison between the numerical estimates of $ \mu_A(1)=\lim_{t\rightarrow \infty} (1/t)\ln\langle e^{- {\cal A}_t}\rangle$ displayed in Fig. \ref{Fig7} and the values of $\mu(1)$ (solid black line) and $\dot {\cal S}_{{J}}$ (dashed red line)  computed from Eqs. (\ref{Eqmuw3}) and (\ref{EqSdotJ}), respectively. One has $\mu(1)=\dot {\cal S}_{{J}}\le 1/Q_0$ for  $\tau\le\tau_{p,1}\approx 7.37$ and $\tau \ge \tau_{p,2}\approx 8.32$, and $\mu(1)=1/Q_0\le\dot {\cal S}_{{J}}$ for $\tau_{p,1}\le \tau\le \tau_{p,2}$. Note that  $\dot {\cal S}_{{\cal J}}$ is a tighter bound to the extracted work rate (dashed-dotted black line) than $1/Q_0$ for $\tau<\tau_{p,1}$ and $\tau>\tau_{p,2}$.}
\end{center}
\end{figure} 

For $\tau_{p,1}\le \tau\le\tau_{p,2}$, the theoretical analysis only indicates that 
\begin{align}
\mu_{\Sigma}(1)\ne \mu(1)=1/Q_0\,.
\end{align}
The latter relation comes from the divergence of the prefactor $g_{\Sigma}(1)$ (cf. Eq. (\ref{Eqgsigma1})). In addition, there is strong evidence from the numerical data displayed  in Fig. \ref{Fig7}  that $\mu_{\Sigma}(1)$ is  equal to $\dot S_{\cal J}$ for {\it all} values of $\tau$, as illustrated by the dashed red line in Fig. {\ref{Fig8}~\cite{note56}. This implies that  $\mu_{\Sigma}(\lambda)$ is discontinuous at $\lambda=1$ for $\tau_{p,1}\le \tau\le\tau_{p,2}$, which  is consistent with the behavior of $(1/t)\ln Z_{\Sigma}(\lambda,t)$ in the vicinity of $\lambda=1$ observed in Fig. \ref{Fig6} for $\tau=7.6$. Similarly, since 
\begin{align}
\mu_Q(1)=1/Q_0\ne \mu(1)=\dot S_{\cal J}\ \ \mbox{for} \ \tau <\tau_{p,1} \ \mbox{and}\  \tau >\tau_{p,2} \,,  
\end{align}
$\mu_Q(\lambda)$ is discontinuous at $\lambda=1$, which is also consistent with the behavior of $(1/t)\ln Z_Q(\lambda,t)$ observed in  Fig. \ref{Fig6} for $\tau=8.4$.

More precisely, inspired by the  exact boundary layer analysis performed in  Appendix B in the small-$\tau$ limit and in the associated Markovian model (see in particular Fig. \ref{FigB1}), we conjecture that 
\begin{align}
\label{EqZQasymp}
Z_{\Sigma}(\lambda,t)e^{-\mu(\lambda) t}&\sim e^{(\dot S_{\cal J}-1/Q_0)t} \ \ \mbox{for} \ \tau_{p,1}<\tau <\tau_{p,2}  \nonumber\\
Z_{Q}(\lambda,t)e^{-\mu(\lambda) t}&\sim e^{(1/Q_0-\dot S_{\cal J})t} \ \ \mbox{for} \ \tau <\tau_{p,1} \ \mbox{and}\  \tau >\tau_{p,2} 
\end{align}
as $t \to \infty$ and $\lambda \to 1$. 
 Clearly, this anomalous behavior of the two SCGFs can be ascribed to the unbounded (but different) growth of the temporal boundary terms, $\Sigma_t-\beta{\cal W}_t=\ln p({\bf x}_i)/p({\bf x}_f)-\beta\Delta {\cal U}({\bf x}_i,{\bf x}_f)$ and  $\beta {\cal Q}_t-\beta {\cal W}_t=\beta\Delta {\cal U}({\bf x}_i,{\bf x}_f)$. In contrast, $\mu_W(\lambda)$ is always equal $\mu(\lambda)$ and is therefore a continuous function of $\lambda$, which is the reason why we have treated ${\cal W}_t$ differently from ${\cal Q}_t$ and $\Sigma_t$ in Sec. III.B.

The analysis performed in Appendix A also gives us some partial information about the values of the prefactors for $\lambda=1$ when these quantities are finite. This is an interesting outcome since, as we have already pointed out, we are unable to compute the prefactors in general. 

For $\tau_{p,1}<\tau<\tau_{p,2}$, after replacing the stationary distributions by their Gaussian expressions in Eq. (\ref{Eqgw1}), we obtain
\begin{align}
\label{Eqgw10}
g_W(1)=\frac{T^2}{[(T-T_x)(T+\hat T_x)]^{1/2}[(T-T_v)(T+\hat T_v)]^{1/2}}\ ,
\end{align}
where $\hat T_x$ and $\hat T_v$ are the  steady-state effective temperatures associated with the hat dynamics. (Recall that $g_Q(1)=1$ and $g_{\Sigma}(1)$ diverges in this case.) 

For $\tau <\tau_{p,1}$ and $ \tau >\tau_{p,2} $, the information is more limited since we cannot compute $g_{W}(1)$ and $g_{\Sigma}(1)$ separately (while $g_Q(1)$ diverges). On the other hand, from Eqs. (\ref{Eqgw2:subeqns}), the ratio of these two prefactors is expected to be
\begin{align}
\label{Eqgw2}
\frac{g_W(1)}{g_{\Sigma}(1)}=\frac{T^2}{[T(T_x+\widetilde T_x)-T_x\widetilde T_x]^{1/2}[T(T_v+\widetilde T_v)-T_v\widetilde T_v]^{1/2}} \ ,
\end{align}
where $\widetilde T_x$ and $\widetilde T_v$ are the  steady-state effective temperatures associated with the tilde dynamics. The variations of $\hat T_v$ and $\widetilde T_v$ with $\tau$ are shown in  Fig. \ref{FigA1}. It is worth noting that $\hat T_v$ and $\widetilde T_v$ are larger than $T_v$ in the stationary cooling regime where $T_v<T$.



\subsubsection{Calculation of $I(w)$}

We now compute the large deviation rate functions and start with $I(w)$. Our basic assumption is that the prefactor $g_W(\lambda)$ has no singularity {\it whatever} the value of $\lambda$ (and not only for $\lambda=1$ as discussed above).
This is supported by the exact analytical calculations in the Markovian limit reported in  Appendix B and is also in line with the exact behavior observed in other (Markovian) nonequilibrium models~\cite{VC2003,BJMS2006,TC2007,KKP2014} and checked experimentally~\cite{B2009,D2006}. Consequently, the LDF $I(w)$ is always given by the Legendre transform   $I(w)=-\lambda^*w-\mu(\lambda^*)$ with $\mu'(\lambda^*)+w=0$. From Eqs. (\ref{Eqmuw1})-(\ref{Hlambda}), this amounts to solving numerically the equation 
\begin{align}
\frac{1}{2\pi}\int_{-\infty}^{\infty}d\omega\: \: \omega \sin(\omega\tau)H_{\lambda^*}(\omega) =-\frac{Q_0^2}{2g}w
\end{align}
so as to obtain the saddle point $\lambda^*$ as a function of $w$. This leads to the curves $e^{-I(w)t}$ shown in Figs. \ref{Fig9} and \ref{Fig10} as dashed black lines.
\begin{figure}[hbt]
\begin{center}
\includegraphics[trim={0 1.25cm 0 1.75cm},clip,width=8cm]{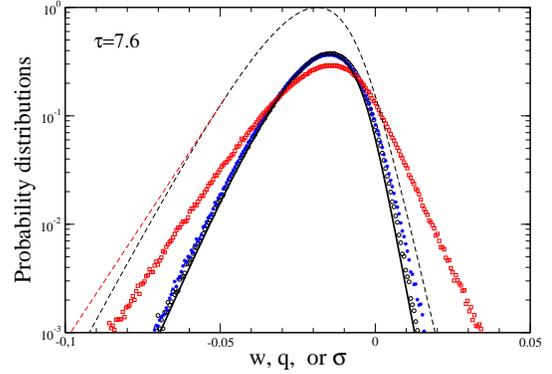}
 \caption{\label{Fig9} (Color on line)  Probability distribution functions $P({\cal A}_t=at)$ for $\tau=7.6$. The symbols are the data obtained from the numerical simulations (see Fig. 4), the dashed black line is the large-deviation form $e^{-I(w)t}$, and the solid black line is the semi-empirical asymptotic expression given by Eq. (\ref{EqPWfull}).  The dashed red line on the l.h.s. for $\sigma\lessapprox -0.048$  is the  curve $e^{-I_1(\sigma)t}$ obtained from Eq. (\ref{EqLDFq}).}
\end{center}
\end{figure}
\begin{figure}[hbt]
\begin{center}
\includegraphics[trim={0 1.25cm 0 1.75cm},clip,width=8cm]{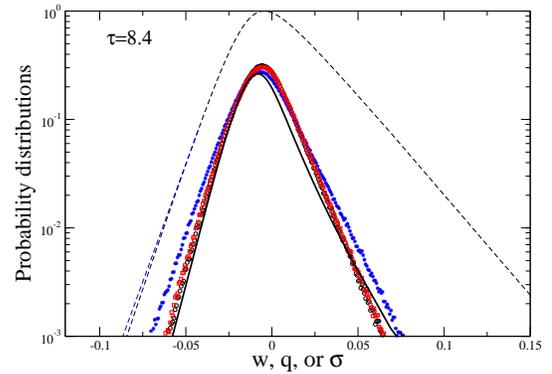}
 \caption{\label{Fig10} (Color on line) Same as Fig. \ref{Fig9} for $\tau=8.4$. The dashed blue line on the l.h.s. for $q\lessapprox -0.042$ is the curve $e^{-I_1(q)t}$ obtained from Eq. (\ref{EqLDFq}).}
\end{center}
\end{figure}

From these figures, however, it is clear  that the large deviation form does not properly describe $P({\cal W}_t=wt)$ for $t=100$, in particular in  Fig. \ref{Fig10} where the slopes on the right-hand side are quite different.  This can be traced back to finite-time corrections which can be  computed by using a standard expansion around the saddle-point (see e.g.~\cite{Sab2011,MR2012}), assuming again the absence of  any singularity in $g_W(\lambda)$. This yields 
\begin{align}
\label{EqPWfull}
P({\cal W}_t=wt)\approx \frac{g_W(\lambda^*(w))}{\sqrt{2\pi \mu''(\lambda^*(w))t}}e^{-I(w)t}\ .
\end{align}
Although the analytical expression of $g_W(\lambda)$  for generic values of $\lambda$ is unknown, a semi-empirical estimate can be obtained from Fig. {\ref{Fig6}, assuming that the very small deviations  between $\mu_W(\lambda)$ and $\mu(\lambda)$ are due to neglecting the prefactor. We thus compute the prefactor as $g_W(\lambda)\approx Z_W(\lambda,t)e^{-\mu(\lambda)t}$, where $Z_W(\lambda,t)$ is obtained from the numerical simulations, and insert the result into Eq. (\ref{EqPWfull})~\cite{note3}. As shown by the solid black lines in Figs.  \ref{Fig9} and \ref{Fig10}, this procedure leads to a much better description of the numerical data. We take this  as an indirect but  convincing evidence that our theoretical analysis of the work fluctuations  is well sounded. The remaining discrepancies observed  for $\tau=8.4$ may be attributed to statistical uncertainty due to the limited sampling.

\subsubsection{Calculation of  $I(q)$ and $I(\sigma)$}

The calculation of the LDFs $I(q)$ and $I(\sigma)$ is more challenging because we can no longer assume that  the prefactors $g_Q(\lambda)$ and $g_{\Sigma}(\lambda)$ have no singularities. In particular, we  already know from the preceding discussion that $\lambda=1$ is a pole of $g_Q(\lambda)$ for $\tau=8.4$ (as $\mu_Q(1)=1/Q_0\ne \mu(1)$)  and a pole of  $g_{\Sigma}(\lambda)$ for $\tau=7.6$ (as $\mu_{\Sigma}(1)=\dot S_{{\cal J}}\ne \mu(1))$.  In addition, the exact calculation of the generating functions $Z_Q(\lambda,t)$ and $Z_{\Sigma}(\lambda,t)$ in the small-$\tau$ limit and in the associated Markovian model shows that other pole singularities appear when performing the stationary average over the {\it initial} state ${\bf x}_i$ (see Eqs. (\ref{EqgA:subeq2}) and (\ref{EqgA:subeq3})). These poles, due again to rare but large fluctuations of the temporal boundary terms, occur for  $\lambda<0$ and  lead to an exponential tail in the r.h.s  of the pdfs~\cite{note12}.  (In contrast,  the poles at $\lambda=1$ occur when performing the average over the {\it final} state ${\bf x}_f$ and lead to an exponential tail in the l.h.s. of the LDFs.)  We then expect that these rare events are  responsible - together with finite-time corrections - for the fact that the slopes of $P(\Sigma_t=\sigma t)$ in Fig.  \ref{Fig9} and of $P({\cal Q}_t=qt)$  in Fig.  \ref{Fig10} are not correctly described by  the Legendre transform of $\mu(\lambda)$.

Unfortunately,  we have no way to determine analytically {\it all} the poles of  $g_Q(\lambda)$ and $g_{\Sigma}(\lambda)$ for an arbitrary value of $\tau$.
The best we can do is to describe how the  pole at $\lambda=1$ (when it exists) modifies the LDFs $I(q)$ and $I(\sigma)$. To this aim, we compute the special value of $q$ or $\sigma$ for which the saddle point $\lambda^*$ reaches $1$. According to Eq. (\ref{Eqmulambda}), it is given by
\begin{align}
\label{Eqastar}
a^*=-\mu'(1)=-\frac{g}{\pi Q_0^2}\int_{-\infty}^{\infty} d\omega\: \omega \sin (\omega \tau) \vert \widetilde\chi(\omega)\vert^2\ ,
\end{align}
where $a^*$ stands for either $q^*$ or $\sigma^*$ and we have used $\widetilde\chi$ in place of $\chi$. When this corresponds to a pole in the prefactor (depending on the observable and on the value of  $\tau$), the LDF becomes linear for $a<a^*$ and is given by
\begin{align}
\label{EqLDFq}
I_1(a)=-[\mu(1)-a]\ .
\end{align}
This leads to the modified asymptotic behaviors $P({\Sigma}_t=\sigma t)\sim e^{-I_1(\sigma) t}$ and $P({\cal Q}_t=q t)\sim e^{-I_1(q)t}$  shown in Figs.  \ref{Fig9}  and \ref{Fig10}, respectively. We see that the slopes on the l.h.s. are now in much better agreement with the numerical simulations. 

\subsubsection{Two stationary-state fluctuation theorems (SSFTs)}

To end our study, we now examine the status of the conventional fluctuation relation for the work ${\cal W}_t$ and state two alternative relations that hold in the long-time limit.
 
\begin{figure}[hbt]
\begin{center}
\includegraphics[width=9cm]{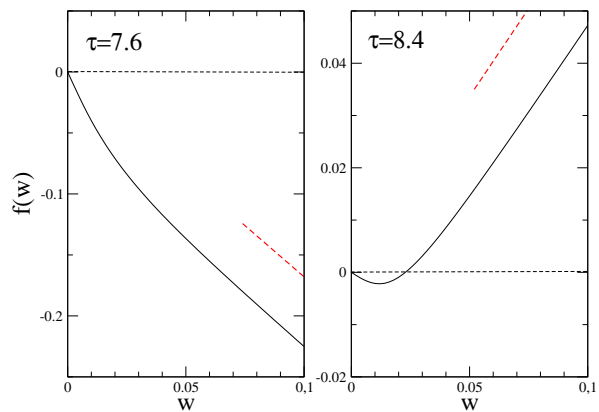}
\caption{\label{Fig11} (Color on line) Symmetry function for the work fluctuations $f(w)=I(-w)-I(w)$. The dashed red lines represent the asymptotic regime of large fluctuations $f(w)\sim (\lambda_{min}+\lambda_{max})w$ (see text).}
\end{center}
\end{figure}

 As we have already mentioned, the SCGF $\mu(\lambda)$, whose  expression is given by Eq. (\ref{Eqmulambda}) or Eq. (\ref{Eqmuw1}), does not possess the symmetry $\mu(1-\lambda)=\mu(\lambda)$ that would lead to a conventional SSFT expressing the symmetry around $0$ of the pdf of an observable ${\cal A}_t$ at large times. 
This is strikingly illustrated by Fig. \ref{Fig11} where we plot the symmetry function $f(w)=I(-w)-I(w)=\lim_{t \to \infty}\frac{1}{t}\ln \frac{P({\cal W}_t=wt)}{P({\cal W}_t=-wt)}$ for $w\ge 0$ (with $f(-w)=-f(w)$). We see that the  SSFT symmetry $f(w)=w$  is violated for all values of $w$. On the one hand, one has $f(w)<0$ for small positive values of $w$ since the average work rate is negative in the cooling regime (as can be seen in Figs. \ref{Fig4} and \ref{Fig5}, the probability of having a negative event $\beta {\cal W}_t=-wt$ is indeed larger than the probability of having a positive event $\beta {\cal W}_t=+wt$).   On the other hand,  large fluctuations are described at the level of the large deviation function by $I(w)\sim -\lambda_{max}w$ for $w<0$ and $I(w)\sim -\lambda_{min}w$ for $w>0$, where $\lambda_{max}$ and $\lambda_{\min}$ are the boundaries of the region of convergence of $\mu(\lambda)$~\cite{T2009} (see Fig. \ref{Fig6} with $\lambda_{\min}\approx -2.94,  \lambda_{\max}\approx 1.26$ for $\tau=7.6$ and $\lambda_{\min}\approx -0.43,  \lambda_{\max}\approx 1.10$ for $\tau=8.4$). This implies that $f(w)\sim (\lambda_{min}+\lambda_{max})w$. As can be seen in Fig. \ref{Fig11}, the symmetry function $f(w)$ smoothly interpolates between these two regimes of small and large fluctuations. The remarkable feature is that the second fluctuation regime is quite different for $\tau=7.6$ and $\tau=8.4$ as $\lambda_{min}+\lambda_{max}<0$ in the first case and $\lambda_{min}+\lambda_{max}>0$ in  the second one.  We emphasize that this striking  effect of the time delay cannot be attributed to the influence of temporal boundary terms since we only focus here on the fluctuations of the work.
 
Whereas the standard symmetry $\mu(1-\lambda)=\mu(\lambda)$ does not hold, it is easily seen from the definition of the function $H_{\lambda}(\omega)$ [Eq. (\ref{Hlambda})] that
\begin{subequations}
\label{EqHlambda:subeqns}
\begin{align}
\hat H_{\lambda-1}(\omega)&=H_{\lambda}(\omega)\label{EqHlambda:subeq1}\\
\widetilde H_{1-\lambda}(\omega)&=H_{\lambda}(\omega)\label{EqHlambda:subeq2}\ ,
\end{align}
\end{subequations}
where $\hat H_{\lambda}(\omega)\equiv H_{\lambda}(\omega)\vert_{\gamma \to -\gamma}$ and $\widetilde H_{\lambda}(\omega)\equiv H_{\lambda}(\omega)\vert_{\tau \to -\tau}$ (we remind the reader that $\hat \chi(\omega)^{-1}= -\omega^2+i\omega/Q_0+1-(g/Q_0)e^{i\omega \tau}$ and $\widetilde \chi(\omega)^{-1}= -\omega^2-i\omega/Q_0+1-(g/Q_0)e^{-i\omega \tau}$ in dimensionless units). We then deduce from Eq. (\ref{Eqmuw1}) the two symmetry relations
\begin{subequations}
\label{Eqmulambda1:subeqns}
\begin{align}
\hat \mu(\lambda-1)&=\mu(\lambda)-\mu(1)\label{Eqmulambda1:subeq1}\\
\widetilde \mu(1-\lambda)&=\mu(\lambda)-\mu(1)\label{Eqmulambda1:subeq2}\ ,
\end{align}
\end{subequations}
where $\hat \mu(\lambda)\equiv \mu(\lambda)\vert_{\gamma \to -\gamma}$ and $\widetilde \mu(\lambda)\equiv\mu(\lambda)\vert_{\tau \to -\tau}$.

Now, for a SSFT to hold, a stationary state must also exist with the  dynamics associated with the transformation $\gamma \to -\gamma$ or $\tau \to -\tau$. In this case, the corresponding pdfs  $\hat P(\beta {\cal W}_t=wt)$ and $\widetilde P(\beta \widetilde{\cal W}_t=wt)$ are expected to acquire asymptotically the large-deviation forms 
\begin{subequations}
\label{EqhPtP:subeqns}
\begin{align}
\hat P(\beta {\cal W}_t=wt)&\sim e^{-\hat I(w)t}\label{EqhPtP:subeq1}\\
\widetilde P(\beta \widetilde{\cal W}_t=wt)&\sim e^{-\widetilde I(w)t}\label{EqhPtP:subeq2}\ ,
\end{align}
\end{subequations}
where $\beta \widetilde{\cal W}_t\equiv \beta {\cal W}_t\vert_{\tau \to -\tau}=(2g)/(Q_0^2)\int_0^t dt'\: x_{t'+\tau}\circ v_{t'}$. Assuming again that boundary terms are irrelevant for the fluctuations of the work at large times, whatever the dynamics, the two LDFs $\hat I(w)$ and $\widetilde I(w)$ are then given by the Legendre transform of the corresponding SCGFs  $\hat \mu(\lambda)$ and $\widetilde \mu(\lambda)$. From Eqs. (\ref{Eqmulambda1:subeqns}) and the corresponding saddle-point equations, we then obtain
\begin{subequations}
\label{EqIw:subeqns}
\begin{align}
I(w)-\hat I(w)&=-w-\mu(1)\label{EqIw:subeq1}\\
I(w)-\widetilde I(-w)&=-w-\mu(1)\label{EqIw:subeq2}\ ,
\end{align}
\end{subequations}
which yields the two SSFTs
\begin{subequations}
\label{EqSSFT:subeqns}
\begin{align}
\lim_{t \to \infty} \frac{1}{t} \ln \frac{P(\beta {\cal W}_t=wt)}{e^{t/Q_0}\hat P(\beta {\cal W}_t=wt)}&=w\label{EqSSFT:subeq1}\\
\lim_{t \to \infty} \frac{1}{t} \ln \frac{P(\beta {\cal W}_t=wt)}{e^{t \dot S_{\cal J}}\widetilde P(\beta \widetilde{\cal W}_t=-wt)}&=w\label{EqSSFT:subeq2}\ .
\end{align}
\end{subequations}
We stress that the fluctuation relation (\ref{EqSSFT:subeq1}) holds for $\tau_{p,1}<\tau<\tau_{p,2}$ in the second stability lobe (hence $\mu(1)=1/Q_0$) whereas relation (\ref{EqSSFT:subeq2}) holds for  $\tau<\tau_{p,1}$ and $\tau>\tau_{p,2}$ (hence $\mu(1)=\dot S_{\cal J}\ne 1/Q_0$).  In fact, since $\widetilde P(\beta \widetilde{\cal W}_t=-wt)\sim \widetilde P(\beta {\cal W}_t=wt)$ asymptotically,  this latter relation can be also re-expressed as
\begin{align}
\label{SSFT2}
\lim_{t \to \infty} \frac{1}{t} \ln \frac{P(\beta {\cal W}_t=wt)}{e^{t \dot S_{\cal J}}\widetilde P(\beta {\cal W}_t=wt)}&=w\ .
\end{align}
\begin{figure}[hbt]
\begin{center}
\includegraphics[width=9cm]{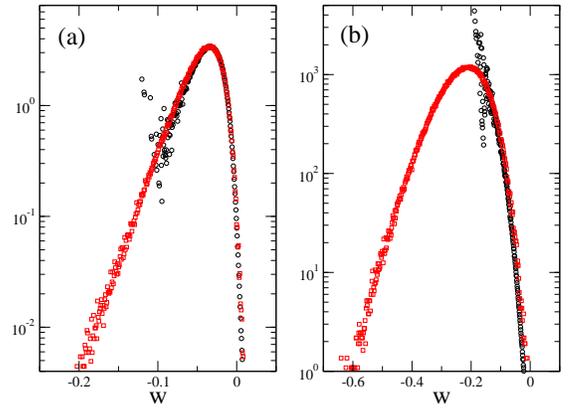}
\caption{\label{Fig12} (Color on line)  Verification of the SSFTs (\ref{EqSSFT:subeq1}) and (\ref{SSFT2}). The figure compares $P(\beta{\cal W}_t=wt)e^{-wt}$ (black circles) with (a) $\hat P({\cal W}_t=wt)e^{t/Q_0}$ (red  squares) for $Q_0=34.2, g/Q_0=0.25,\tau=7.6$, and (b) $\widetilde P(\beta{\cal W}_t=wt)e^{\dot {\cal S}_{\cal J}t}$ (red squares) for $Q_0=2, g/Q_0=0.55,\tau=2.5$ ($\dot S_{\cal J}\approx 0.1$).  The observation time is $t=100$.}
\end{center}
\end{figure}

A numerical check of the two SSFTS is provided  in Figs. \ref{Fig12}(a)   and  \ref{Fig12}(b) (see below for an explanation of the numerical procedure). The agreement is  satisfactory in both cases, taking into account that the exponential factor $e^{-\beta {\cal W}_t}$ strongly weights  work values  in the far left tail of $P(\beta{\cal W}_t)$ corresponding to very rare realizations of the process that cannot be properly sampled~\cite{J2006}. As $t$ increases, we expect the curves in Fig. \ref{Fig12}  to be peaked more and more around the asymptotic work value $w^*=\lim_{t \to \infty}\int dw \:w P(\beta{\cal W}_t)e^{-wt}/\int dw P(\beta{\cal W}_t)e^{-wt}=-\mu'(1)$ (cf. Eq. (\ref{Eqastar})), with $w^*\approx -0.048$ in Fig. \ref{Fig12}(a)  and $w^*\approx-0.236$ in Fig. \ref{Fig12}(b).  This latter figure illustrates the curious feature that atypical fluctuations become typical when generated by an acausal dynamics ! This dynamics (when it leads to a stationary state) then defines the so-called  ``auxiliary" or ``driven" process~\cite{GL2010,JS2010,CT2013} that generates asymptotically the ensemble of paths conditioned on the  constraint $\beta {\cal W}_t/t=w^*$~\cite{note67}. In fact, changing $\tau$ into $-\tau$ in the stationary cooling regime $T_v<T$ has essentially the same effect as changing $\gamma$ into $-\gamma$, namely to {\it enhance} the fluctuations in the system and thus to increase the effective temperatures: for instance,  one has $\hat T_x\approx 0.792>T_x\approx 0.420$, $\hat T_v\approx 0.656> T_v\approx 0.359$, and $\widetilde T_x\approx 1.538>T_x\approx0.956$, $\widetilde T_v\approx 1.220> T_v\approx0.960$  for the two cases represented in Fig. \ref{Fig12} (see also Fig. \ref{FigA1}). But, at the same time,  there is {\it more} work extracted from the bath since in both cases $w^*$  is more negative than the average work - or dissipated heat- rate $(1/Q_0)(T_v/T-1)$ (cf. Eq. (\ref{Eqavheat})). This kind of counterintuitive behavior that occurs in the rare fluctuations regime is discussed in Ref. \cite{RH2016} for another model of feedback cooling, where the focus is on the information exchange between the system and the feedback controller. 

It is instructive to detail how the numerical data displayed in Fig. \ref{Fig12}(b) were obtained. For the $\gamma \to -\gamma$ dynamics, one can directly solve the dimensionless Langevin equation $\dot v_t=(1/Q_0)v_t -x_t+(g/Q_0) x_{t-\tau}+\xi_t$ using the standard  Euler or Heun's methods. However these schemes cannot be applied to the acausal  Langevin equation $\dot v_t=-(1/Q_0)v_t -x_t+(g/Q_0) x_{t+\tau}+\xi_t$. Fortunately, thanks to the linearity of the equation, there is a strategy for tackling this problem. Indeed, for a given history of the thermal noise $\xi(t)$ over a long time interval $[-t_1,t_2]$, a stationary solution  can be approximated as
\begin{align}
\label{Eqxttilde}
 x(t)\approx\int_{-t_1}^{t_2} dt' \widetilde \chi(t-t) \xi(t')\ ,
\end{align}
where $\widetilde \chi(t)$ is the  inverse Fourier  transform of the acausal response function $\widetilde \chi(\omega)$ (see the discussion in Appendix A). If  $t_1,t_2\gg t >0$ and if $\widetilde \chi(t)$ decays sufficiently rapidly for both positive and negative times, Eq. (\ref{Eqxttilde})  provides a very good approximation of $x(t)$ in the time interval $[0,t]$. In this way, one can generate a representative ensemble of stationary trajectories and estimate the probabilities $\widetilde P(\beta \widetilde{\cal W}_t=-wt)$ or $\widetilde P(\beta {\cal W}_t=wt)$. (It turns out that the case $Q_0 = 34.2,g/Q_0=0.25,\tau=8.4$ cannot be studied with this method because $\widetilde \chi(t)$ decays too slowly for $t>0$ as the two poles of $\widetilde \chi(s)$ on the l.h.s. of the complex $s$-plane have a very small real part. Fig. \ref{Fig12}(b) thus corresponds to another choice of the parameters for which a stationary state still exists with the acausal dynamics and $\widetilde \chi(t)$ decays to $0$ rapidly,  as shown in Fig. \ref{Fig13}.)
\begin{figure}[hbt]
\begin{center}
\includegraphics[width=7cm]{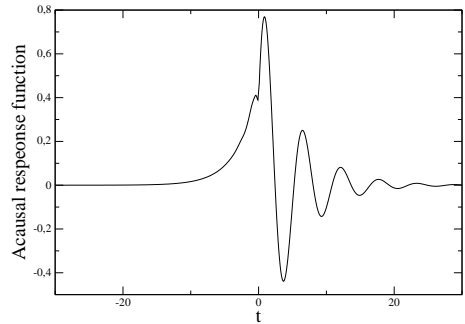}
\caption{\label{Fig13}  Acausal response function $\widetilde{\chi}(t)$ for $Q_0=2, g/Q_0=0.55$ and $\tau=2.5$. $\widetilde \chi(s)$ has two poles on the l.h.s. of the complex $s$-plane and an infinite number of poles on the r.h.s. The poles $s\approx -0.200\pm1.12i$ and $s\approx 0.339 $  control the behavior of $\widetilde \chi(t)$ for $t\ge 0$ and for $t \to -\infty$, respectively.}
\end{center}
\end{figure}

Finally, we mention another way to understand the origin of the large fluctuations contributing to  ${\cal P}(\beta {\cal W}_t)e^{-\beta {\cal W}_t}$, which is to consider the atypical thermal noise that generates such fluctuations. To this end, we select an atypical stationary trajectory produced by one or the other  conjugate process and insert it into the original Langevin equation. The calculation can be readily performed in  the frequency domain, which yields, for instance in the case of the acausal dynamics, 
\begin{align}
\label{Eqatyp}
\xi_{atyp}(\omega)=\frac{\widetilde \chi(\omega) }{\chi(\omega) }\xi(\omega)\ . 
\end{align}
The atypical  noise is thus  colored, with autocorrelation function $\langle \xi_{atyp}(t)\xi_{atyp}(t')\rangle=\nu(t-t')$ given by
\begin{align}
\nu(t)&= \delta(t)+\frac{1}{2\pi}\int_{-\infty}^{+\infty}d\omega\: (\frac{\vert\widetilde \chi(\omega)\vert^2}{\vert\chi(\omega)\vert^2}-1)e^{-i\omega t}\nonumber\\
&= \delta(t)+\frac{2g}{\pi Q_0^2}\int_{-\infty}^{+\infty}d\omega \:\omega \sin(\omega \tau) \vert \widetilde \chi(\omega)\vert^2e^{-i\omega t}
\end{align}
 in dimensionless  units. An illustration is provided in Fig. \ref{Fig14} for the same model parameters used in Fig. \ref{Fig12}(b) and Fig. \ref{Fig13}.
 \begin{figure}[hbt]
\begin{center}
\includegraphics[width=7cm]{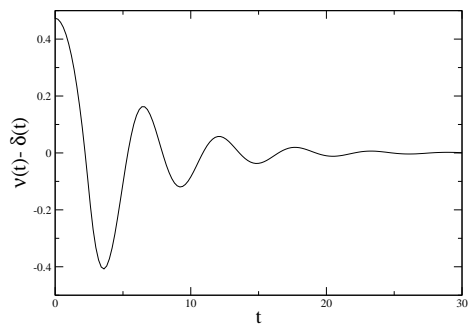}
\caption{\label{Fig14} (Color on line) Autocorrelation function of the atypical colored noise $\xi_{atyp}(t)$  for $Q_0=2, g/Q_0=0.55$ and $\tau=2.5$.}
\end{center}
\end{figure}

\section{Summary and closing remarks}

In this paper we have investigated the nonequilibrium steady-state fluctuations of thermodynamic observables in a Brownian system subjected to a time-delayed feedback control, focusing on the behavior at large times. Our study, based on both analytical and numerical calculations, has revealed that the delay significantly affects the large-deviation statistics of time-integrated thermodynamic observables. In particular, when the state space is unbounded, delay plays a critical role in the occurrence of rare but large fluctuations of temporal boundary terms so that observables with the same typical value exhibit different large deviation rate functions.

Compared to the Markovian case, there is no doubt that the study of time-delayed systems presents some new challenges. From the perspective of stochastic thermodynamics, the most delicate issue is that the behavior of the system under time reversal is modified, which prevents standard fluctuation theorems from being satisfied. Hidden symmetries do exist but  their interpretation is more subtle, as shown in this work, and a complicated analysis of the response function of the conjugate dynamics is required even in the simplest case of a linear dynamics. In fact, it is remarkable that the large-deviation statistics, which in principle is accessible to experiments, cannot be fully elucidated without investigating the unusual properties of an acausal dynamics. Taking into account the ubiquity of time-delayed feedback loops in natural and artificial systems, there is obviously an avenue for future investigations.
 

\appendix 
\setcounter{figure}{0} \renewcommand{\thefigure}{A.\arabic{figure}} 

\renewcommand{\theequation}{A\arabic{equation}}

\section{Conjugate dynamics and asymptotic behavior of $\langle e^{-\beta {\cal W}_t}\rangle$ and $\langle e^{-\Sigma_t}\rangle$}

In this Appendix, we show how Eqs. (\ref{EqZA3}) and (\ref{EqZA4}) in the main text can be used  to infer the  long-time behavior of the generating functions $Z_W(\lambda,t)$ and $Z_{\Sigma}(\lambda,t)$ for  $\lambda=1$. (On the other hand, we know from Eq. (\ref{EqIFTQ}) that $Z_Q(1,t)= e^{(\gamma/m)t}$ at all times.) For concreteness, we restrict the discussion to the case of the linear Langevin equation (\ref{EqL}) considered in Sec. IV. The following equations are thus expressed in terms of  dimensionless parameters. For instance, the exponential factor $e^{(\gamma/m)t}$  becomes $e^{t/Q_0}$. A similar analysis has been performed in Ref. \cite{RTM2016} in the context of heat flow in harmonic chains. 

\subsection{Stationary solutions of the conjugate dynamics}

The first task is to determine under which conditions a stationary solution of the conjugate Langevin equations Eq. (\ref{EqL1minus}) and Eq. (\ref{EqLtilde}) exists. (As usual, a solution is called stationary if the $n$-point probability distributions are invariant under time translation.) In the first case of the so-called ``hat" dynamics, the existence of a stationary state means that an arbitrary initial condition is eventually forgotten, i.e.,
\begin{align}
x(t)\approx \int_{-\infty}^tdt'\: \hat \chi(t-t')\xi(t')\ ,
\end{align}  
with the response function $\hat \chi(t)$ decreasing sufficiently fast (typically exponentially) for $t \to +\infty$.  The function $\hat \chi(\omega)\equiv \chi(\omega)\vert_{\gamma \to -\gamma}$  is then the genuine Fourier transform of $\hat \chi(t)$, i.e., $\hat \chi(\omega)=\int_{-\infty}^{\infty} dt\: e^{i\omega t}\hat \chi(t)$. Since $\hat \chi(t)$ is causal, this requires that all the poles of $\hat \chi(\omega)$ lie in the lower half of the complex $\omega$-plane (equivalently, all the poles of  $\hat \chi(s)=\int_{-\infty}^{\infty} dt\: e^{-s t}\hat \chi(t)$  lie in the l.h.s. of the complex Laplace plane $s=\sigma-i\omega$). 

The  case of Eq. (\ref{EqLtilde}) is more subtle because the so-called ``tilde" dynamics is acausal. The stationary state, if it exists,  must then  be  independent of both the initial condition in the far past {\it and} the final condition in the far future. Although this may seem an awkward requirement, this simply means that 
\begin{align}
x(t)\approx\int_{-\infty}^{+\infty}dt'\: \widetilde\chi(t-t')\xi(t')\ ,
\end{align} 
with the acausal response function $\widetilde \chi(t)$ decreasing sufficiently fast for both  $t \to +\infty$  and $t \to -\infty$ (see Fig. \ref{Fig13} in the main text). Then, $\widetilde \chi(\omega)\equiv \chi(\omega)\vert_{\tau \to -\tau}$ is the Fourier transform of $\widetilde \chi(t)$, and conversely. However, as explained in I (see Eq. (161) and appendix E),  $\widetilde \chi(t)$ is more generally  defined as the inverse bilateral Laplace transform of $\widetilde \chi(s)$, i.e., $\widetilde \chi(t)=1/(2\pi i)=\int_{c-i\infty}^{c+i\infty} dt\: e^{st}\widetilde \chi(s)$, with the {\it same} Bromwich contour $\mbox{Re}(s) = c$ as the one used for computing the quantity $\dot S_{\cal J}$. Therefore, for $\widetilde \chi(t)$ to be the inverse  Fourier transform of $\widetilde \chi(\omega=is)$, which corresponds to $c=0$,  the bilateral Laplace transform $\widetilde \chi(s)$ must have two and only two poles on the l.h.s. of the complex $s$-plane. (In contrast, the functions $\widetilde \chi(t)$ plotted in Figs. 18 and 19 of I have no Fourier transform.)

Since $\hat \chi(s)=\widetilde \chi(-s)$, which is a consequence of the  general relation (\ref{Eqactionrelation}) between the OM actions $\hat S[{\bf X},{\bf Y}]$ and $\widetilde S[{\bf X}^\dag,{\bf Y}^\dag]$, we may re-phrase the conditions for the existence of a stationary state as follows:  A stationary solution of Eq. (\ref{EqL1minus}) exists when all the poles of $\widetilde \chi(s)$ lie in the r.h.s of the complex $s$-plane (case 1), and a stationary solution of Eq. (\ref{EqLtilde}) exists when two and only two poles of $\widetilde \chi(s)$ lie in the l.h.s. (case 2). 
\begin{figure}[hbt]
\begin{center}
\includegraphics[trim={0 1.25cm 0 1.75cm},clip,width=8cm]{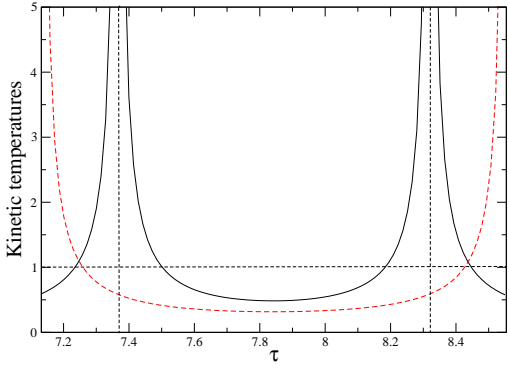}
\caption{\label{FigA1} (Color on line) Kinetic temperatures computed from Eq. (\ref{EqhatT:subeq2}): the black solid line is $\hat T_v/T$ for $7.37<\tau<8.32$ and $\widetilde T_v/T$ for $\tau<7.37$ and $\tau >8.32$. The dashed red line is the kinetic temperature $T_v/T$ of the original dynamics which diverges at the boundaries of the stability region.}
\end{center}
\end{figure}

The stationary distributions $\hat p(x,v)$ and $\widetilde p(x,v)$, when they exist, are  bivariate Gaussians  characterized by the variances of $x$ and $v$ or, equivalently, by the corresponding effective temperatures which we denote by $\hat T_x, \hat T_v$ and  $\widetilde T_x, \widetilde T_v$, respectively. By definition, the variances are obtained by integrating the power spectral density over frequency. Therefore, since $\vert \hat \chi(\omega)\vert^2=\vert \widetilde\chi(\omega)\vert^2$, the temperatures $\hat T_x$ and $\widetilde T_x$ (resp. $\hat T_v$ and $\widetilde T_v$) are given by the same formulas, i.e.,  in terms of dimensionless parameters,
\begin{subequations}
\label{EqhatT:subeqns}
\begin{align}
\frac{2}{Q_0} \int_{-\infty}^{\infty} \frac{d\omega}{2\pi} \: \vert \hat \chi(\omega)\vert^2&=\begin{cases} \hat T_x/T \ , \mbox{ in case 1}\\ \widetilde T_x/T \ , \mbox{ in case 2}\end{cases} \label{EqhatT:subeq1}\\
\frac{2}{Q_0} \int_{\infty}^{\infty} \frac{d\omega}{2\pi}  \: \omega^2\vert \hat \chi(\omega)\vert^2&=\begin{cases} \hat T_v/T\ , \mbox{ in case 1}\\ \widetilde T_v/T\ , \mbox{ in case 2} \end{cases}\label{EqhatT:subeq2}
\end{align}  
\end{subequations} 
We stress, however, that it is only for $\hat T_x$ and $\hat T_v$ that one can repeat the calculation performed in Appendix B of I and obtain closed-form expressions by solving the linear differential equation obeyed by the stationary time-correlation function $\hat \phi(t_2-t_1)=\langle x(t_1)x(t_2)\rangle$ for $0\le \vert t_2-t_1\vert \le \tau$. The expressions of $\hat T_x$ and $\hat T_v$ are  then simply obtained by changing $\gamma$ into $-\gamma$ in  Eqs. (113)-(114) of I.  One can check that this is in agreement with the numerical integration of Eqs. (\ref{EqhatT:subeqns}) {\it only} when the stationary state exists, so that $\langle x(t_1)x(t_2)\rangle$  only depends on $t_2-t_1$ and the calculation in Appendix B of I is applicable. Otherwise, one finds negative temperatures.

 As an illustration, we plot in Fig. \ref{FigA1} the  kinetic temperatures $\hat T_v$ and $\widetilde T_v$  computed  for $Q_0=34.2$, $g/Q_0=0.25$, and when the system operates in the second stability lobe. As predicted by the location of the poles of $\widetilde \chi(s)$, a stationary state exists with the hat dynamics for $7.37<\tau<8.32$ and with the tilde dynamics for $\tau<7.37$ or $\tau >8.32$. 

\subsection{Asymptotic behavior of $\langle e^{-\beta {\cal W}_t}\rangle$ and $\langle e^{-\Sigma_t}\rangle$}

We now use the preceding results to predict the long-time behavior of  $Z_W(1,t)=\langle e^{-\beta {\cal W}_t}\rangle$ and $Z_{\Sigma}(1,t)=\langle e^{-\Sigma_t}\rangle$  from Eqs. (\ref{EqZA3}) and (\ref{EqZA4}).

\subsubsection{``Hat" dynamics}

We first consider the $\gamma \to -\gamma$ ``hat" dynamics and set $\lambda=1$ in Eqs. (\ref{EqKhat})-(\ref{Eq_fA_lambda3:subeqns}). Then, 
\begin{align}
\label{EqKhat1}
\hat{\cal K}_{1}[{\bf x}_f,t\vert {\bf Y}]=e^{t/Q_0}\int_{{\bf x}_i}^{{\bf x}_f}{\cal D}{\bf X} \: \hat {\cal P}[{\bf X}\vert {\bf Y}]\equiv e^{t/Q_0}\: \hat p({\bf x}_f,t\vert {\bf Y})\ ,
\end{align}
where $\hat p({\bf x}_f,t\vert {\bf Y})$ may be viewed as a generalized transition probability (if the ``hat" process were Markovian, $\hat p({\bf x}_f,t\vert {\bf Y})$ would be the standard transition probability $\hat p({\bf x}_f,t\vert {\bf x}_i,0)$). Since $\hat f_{Q,1}({\bf x}_i,{\bf x}_f)=1$, we thus have  $\int d{\bf x}_f \:\hat{\cal K}_{1}[{\bf x}_f,t\vert {\bf Y}]=e^{t/Q_0}\:  \int d{\bf x}_f \:\hat p({\bf x}_f,t\vert {\bf Y})=e^{t/Q_0}$ in Eq. (\ref{EqZA3}), and using $ \int d\mathbb{P}[{\bf Y}]=p({\bf x}_i)$ we  recover the IFT (\ref{EqIFTQ}), as it must be.

We now assume that the conjugate Langevin equation (\ref{EqL1minus})  admits a stationary solution, as discussed above. Initial conditions are then irrelevant  in the long-time limit, so that
\begin{align}
\lim_{t\to \infty}\hat p({\bf x}_f,t\vert {\bf Y})=\hat p({\bf x}_f)\ ,
\end{align}
where $\hat p({\bf x})$ is the corresponding stationary pdf. Eqs. (\ref{EqZA3}) and (\ref{Eq_fA_lambda3:subeqns1})  then lead to the asymptotic expression
\begin{align}
Z_W(1,t)\sim e^{t/Q_0} \int  d{\bf x}_i \: p({\bf x}_i) e^{\beta {\cal U}({\bf x}_i)}\int d{\bf x}_f\: \hat p({\bf x}_f) e^{-\beta {\cal U}({\bf x}_f)}\ ,
\end{align}
which allows us to conclude that
 \begin{align}
\label{EqmuW1}
\mu_W(1)=\frac{1}{Q_0}\ ,
\end{align} 
and 
\begin{align}
\label{Eqgw1}
g_W(1)=\int  d{\bf x}_i \: p({\bf x}_i) e^{\beta {\cal U}({\bf x}_i)}\int d{\bf x}_f\: \hat p({\bf x}_f) e^{-\beta {\cal U}({\bf x}_f)}\ . 
\end{align}
The prefactor is indeed finite as can be checked explicitly by inserting the expression (\ref{Eqpdf1}) of $p({\bf x})$ and the corresponding expression of $\hat p({\bf x})$ (with $T_x$ and $T_v$ replaced by $\hat T_x$ and $\hat T_v$), and performing the integrations over ${\bf x}_i$ and ${\bf x}_f$. This yields Eq. (\ref{Eqgw10}) in the main text, showing that $0<g_W(1)<\infty$ as long as the temperatures $\hat T_x$ and $\hat T_v$ are positive. 

On the other hand, from Eqs. (\ref{EqZA3}) and (\ref{Eq_fA_lambda3:subeqns3}), we obtain
\begin{align}
\label{Eqgsigma1}
Z_{\Sigma}(1,t)\sim e^{t/Q_0} \int  d{\bf x}_i \int d{\bf x}_f \: p({\bf x}_f) \hat p({\bf x}_f) \ ,
\end{align} 
which  shows that the  prefactor diverges. From this, we conclude that  $\mu_{\Sigma}(1)\ne 1/Q_0$, but, unfortunately, we cannot infer the exact value.

\subsubsection{``Tilde" dynamics}

We now turn our attention to the $\tau \to -\tau$ ``tilde"  dynamics (\ref{EqLtilde}). Thanks to the linearity of the Langevin equation, the Jacobian $\widetilde{\cal J}[{\bf X}]$ is  path independent, and setting $\lambda=1$ in Eq. (\ref{EqKtilde}) yields
\begin{align}
\label{EqKtilde1}
\widetilde{\cal K}_{1}({\bf x}_f^\dag,t\vert {\bf x}_i^\dag; {\bf Y}^\dag)&= \frac{{\cal J}_t}{\widetilde {\cal J}_t} \int_{{\bf x}_i^\dag}^{{\bf x}_f^\dag} {\cal D}{\bf X}^\dag\: \widetilde {\cal P}[{\bf X}^\dag\vert {\bf x}_i^\dag,{\bf Y}^\dag]\ .
\end{align}
At first sight, this resembles Eq. (\ref{EqKhat1}), with the ratio ${\cal J}_t/\widetilde {\cal J}_t$ replacing the exponential factor $e^{t/Q_0}$. There are two features, however, that complicate the asymptotic analysis. The first one is that while we know that ${\cal J}_t/\widetilde {\cal J}_t$ grows exponentially as $e^{\dot S_{{\cal J}}t}$, with $\dot S_{{\cal J}}$ given by Eq. (\ref{EqSdotJ}), we do not know the prefactor. The second one is that ${\bf Y}^\dag$ is a trajectory in the time interval $[t,t+\tau]$ (see Fig. \ref{Fig1}). Therefore, even when the system  relaxes toward a stationary state with the tilde dynamics, $\widetilde{\cal K}_{1}({\bf x}_f^\dag,t\vert {\bf x}_i^\dag; {\bf Y}^\dag)$ still depends on ${\bf Y}^\dag$ in the long-time limit and only the  dependence on ${\bf x}_i^\dag$ is lost. Then, asymptotically, the quantity $\int d\mathbb{P}[{\bf Y}]  \int_{{\bf x}_i^\dag}^{{\bf x}_f^\dag} {\cal D}{\bf X}^\dag\: \widetilde {\cal P}[{\bf X}^\dag\vert {\bf x}_i^\dag,{\bf Y}^\dag]$ involves steady-state trajectories ${\bf X}^\dag$ generated by the ``tilde"  dynamics ending at ${\bf x}_f^\dag=(x_i,-v_i)$ and steady-state trajectories ${\bf Y}$ generated by the direct dynamics ending at ${\bf x}_i=(x_i,v_i)$. The only dependence is on $x_i$ and $v_i$, and one expects
\begin{align}
\label{EqK1asymp}
\lim_{t\to \infty} \int d\mathbb{P}[{\bf Y}]  \int_{{\bf x}_i^\dag}^{{\bf x}_f^\dag} {\cal D}{\bf X}^\dag\: \widetilde {\cal P}[{\bf X}^\dag\vert {\bf x}_i^\dag,{\bf Y}^\dag] \propto p({\bf x}_i)\widetilde p({\bf x}_f^\dag)\ ,
\end{align}
where $\widetilde p({\bf x})$ is the stationary pdf of the ``tilde" dynamics. The proportionality factor could in principle depend on $x_i,v_i$. However, in the small-$\tau$ limit and the associated Markovian model (see below), this factor is simply equal to $1$. In the non-Markovian case, and in the overdamped limit which is simpler to analyze (see e.g. Appendix A in I), we have also performed an exact perturbative calculation at the second order in the amplitude of the feedback force. The outcome is again that the prefactor is constant~\cite{note159}. We therefore consider as most plausible that this is the generic behavior.

As a result, we predict the following asymptotic behavior:
\begin{align}
\label{EqK1asymp}
\int d\mathbb{P}[{\bf Y}]  \widetilde{\cal K}_{1}({\bf x}_f^\dag,t\vert {\bf x}_i^\dag; {\bf Y}^\dag)\sim \kappa\: e^{\dot S_{{\cal J}}t} p({\bf x}_i)\widetilde p({\bf x}_f^\dag)\ ,
\end{align}
where  $\kappa$ is some constant depending on the model parameters for which we have no  expression.
Fortunately,  this is sufficient to infer the asymptotic behavior of $Z_W(1,t)$ and $Z_{\Sigma}(1,t)$. Indeed, from Eq. (\ref{EqZA4}) we obtain
\begin{align}
\label{EqZA5}
Z_A(1,t)\sim \kappa e^{\dot S_{{\cal J}}t} \int d{\bf x}_i  \: p({\bf x}_i)\widetilde p({\bf x}_i)\int d{\bf x}_f\:\widetilde f_A({\bf x}_i,{\bf x}_f)\ ,
\end{align}
where  we have used the fact that $p({\bf x})$ and $\widetilde p({\bf x})$ are even function of $v$ to replace ${\bf x}_i^\dag$ and ${\bf x}_f^\dag$ by ${\bf x}_f$ and ${\bf x}_i$, respectively. We deduce that
\begin{align}
\label{Eqmuw4}
\mu_W(1)=\mu_{\Sigma}(1)=\mu(1)=\dot S_{{\cal J}}\ .
\end{align}
and
\begin{subequations}
\label{Eqgw2:subeqns}
\begin{align}
g_W(1)&= \kappa \int  d{\bf x}_i \: p({\bf x}_i) \tilde p({\bf x}_i)e^{\beta {\cal U}({\bf x}_i)}\int d{\bf x}_f\:e^{-\beta {\cal U}({\bf x}_f)}\label{Eqgw2:subeqs1}\\
g_{\Sigma}(1)&=\kappa \int  d{\bf x}_i \: \tilde p({\bf x}_i)\int d{\bf x}_f\: p({\bf x}_f)= \kappa \label{Eqgw2:subeqs2}\ .
\end{align}
\end{subequations}
Interestingly, the unknown factor $\kappa$ cancels out in the ratio $ g_W(1)/g_{\Sigma}(1)$, which yields Eq. (\ref{Eqgw2}) in the main text.
In line with the considerations above, one can check that Eqs. (\ref{Eqmuw4}) and (\ref{Eqgw2:subeqns}) are in agreement with the exact results in the Markovian small-$\tau$ limit for $\gamma>\gamma'$, with  $\dot S_{{\cal J}}=\gamma'/m$, $g_W(1)=1-(\gamma'/\gamma)^2$, and $g_{\Sigma}(1)=1$ (see Appendix B.2), as well as in the perturbative calculation for the overdamped limit of the non-Markovian case~\cite{note159}.

On the other hand,  Eq. (\ref{EqZA5}) yields
\begin{align}
Z_Q(1,t)\sim \kappa e^{\dot S_{{\cal J}}t} \int d{\bf x}_i  \: p({\bf x}_i)\widetilde p({\bf x}_i)\int d{\bf x}_f\ ,
\end{align}
so that the prefactor diverges. This is expected since $\mu_Q(1)=1/Q_0\ne\dot S_{{\cal J}}$ when  two poles of $\widetilde \chi(s)$ lie on the left-hand side of the complex $s$-plane.

\setcounter{figure}{0} \renewcommand{\thefigure}{B.\arabic{figure}} 
\renewcommand{\theequation}{B\arabic{equation}} 

\section{Small-$\tau$ limit and Markovian model}

In order to better understand the stationary-state fluctuations in the feedback-cooling model studied in Sec. IV, it is very useful to investigate in detail the Markovian limit  obtained by expanding the feedback force $F_{fb}(t)=k'x(t-\tau)$ at first order in $\tau$. The Langevin equation (\ref{EqL}) then reads
\begin{align}
\label{EqMR}
m\dot v_t=-(\gamma +\gamma' )v_t -\bar k x_t+\sqrt{2\gamma T}\xi_t \ ,
\end{align}
where $\bar k=k-k'$ and $\gamma'= k' \tau$. This is precisely the model studied in Refs.~\cite{KQ2004,KQ2007,MR2012} whose main characteristic is the dependence of the feedback force $F_{fb}(t)=-\gamma' v_t$ on the particle's velocity. Interestingly,  this induces features that are similar to those encountered in the original non-Markovian  model. The bonus is that  the generating functions $Z_A(\lambda,t)$ in the steady state can be computed exactly at {\it all} times, as shown in this Appendix  that revisits and extends  earlier work by two of us~\cite{MR2012}. (Accordingly,  to be in line with Ref.~\cite{MR2012}, we choose to work with Eq. (\ref{EqMR}) instead of the dimensionless version.)  In passing, we recall that Eq. (\ref{EqMR}) also describes a Brownian particle coupled to two thermostats at temperatures $T$ and $T'$ in the limit $T' \to 0$. The quantity of interest in this model is the heat exchanged between the two baths, and the full expression of $Z_Q(\lambda,t)$ for $T'>0$ was computed in \cite{V2006} in the case of a free Brownian particle, i.e., for $\bar k=0$ (see also Ref.~\cite{Fa2002}). For $\bar k> 0$,  $Z_Q(\lambda,t)$ is only known in the  long-time limit~\cite{KSD2011,SD2011,FI2012}.

In \cite{MR2012}, only the generating function $Z_{\Sigma}(\lambda,t)$ of  the  entropy production functional $\Sigma_t[{\bf X}]= \beta {\cal Q}_t[{\bf X}] +\ln p({\bf x}_i)/p({\bf x}_f)$ was considered (more precisely, it was the generating function of  $\Sigma_t[{\bf X}]+(\gamma'/m)t$, which is the quantity called $\Delta S_p$ in ~\cite{KQ2007}). Here we generalize this calculation  to also include $Z_W(\lambda,t)$ and $Z_Q(\lambda,t)$. In particular, we wish to bring to light some features that were not discussed in ~\cite{MR2012} and that are also relevant to the non-Markovian case. Note that in what follows we consider the Markovian model (\ref{EqMR}) in its full generality, i.e., with no constraints on $\bar k$ and $\gamma'$ (except that they are both positive). The small-$\tau$ limit of Eq. (\ref{EqL}) then corresponds to a restricted range of these parameters.

\subsection{General expression of the generating functions}

The starting point is the path-integral representation of $Z_{A}(\lambda,t)$, Eq. (\ref{EqZA1}), where the dependence on ${\bf Y}$ is replaced by a  dependence on the initial state ${\bf x}_i$ of the trajectory ${\bf X}$. This equation becomes 
\begin{align}
\label{EqZAexact}
Z_A(\lambda,t)=\int d{\bf x}_i  \: p({\bf x}_i)\int d{\bf x}_f \: f_{A,\lambda}({\bf x}_i,{\bf x}_f) {\cal K}_{\lambda}({\bf x}_f,t\vert {\bf x}_i,0)\ ,
\end{align}
where the functions $f_{A,\lambda}({\bf x}_i,{\bf x}_f)$ are defined in Eqs. (\ref{Eq_fA_lambda:subeqns}) and
\begin{align}
\label{EqKexact}
{\cal K}_{\lambda}({\bf x}_f,t\vert {\bf x}_i,0)&=\int_{{\bf x}_i}^{{\bf x}_f} {\cal D}{\bf X}\:  e^{-\lambda \beta {\cal W}_t}{\cal P}[{\bf X}\vert {\bf x}_i]\ ,
\end{align}
with ${\cal W}_t[{\bf X}]=-\gamma'  \int_0^t dt'\:v_{t'}^2$. Since the effective damping constant in Eq. (\ref{EqMR}) is  $\gamma +\gamma'$,  the path probability ${\cal P}[{\bf X}\vert {\bf x}_i]$ can be expressed as 
\begin{align}
{\cal P}[{\bf X}\vert {\bf x}_i]\propto e^{\frac{\gamma +\gamma'}{2m}t}  e^{-\beta {\cal S}[{\bf X}]}
\end{align}
(see Eq. (\ref{Path:subeq1})), where
\begin{align}
{\cal S}[{\bf X}]= \frac{1}{4\gamma} \int_0^t dt'\:[m\dot v_{t'}+(\gamma+\gamma') v_{t'} +\bar k x_{t'}]^2\ . 
\end{align}
Hence
\begin{align}
\label{EqKexact}
{\cal K}_{\lambda}({\bf x}_f,t\vert {\bf x}_i,0)\propto e^{\frac{\gamma +\gamma'}{2m}t} \int_{{\bf x}_i}^{{\bf x}_f}{\cal D}{\bf X} \: e^{-\beta {\cal S}_{\lambda}[{\bf X}]}\ ,
\end{align}
where  
\begin{align}
\label{EqSlambdaM}
{\cal S}_{\lambda}[{\bf X}]&\equiv {\cal S}[{\bf X}]-\lambda\gamma'  \int_0^t dt'\:v_{t'}^2\ .
\end{align}
The crucial feature that distinguishes the small-$\tau$ limit and the associated Markovian model from the full non-Markovian model is that ${\cal S}_{\lambda}[{\bf X}]$  can be written as an  Onsager-Machlup (OM) action functional for {\it all} values of $\lambda$. The function ${\cal K}_{\lambda}({\bf x}_f,t\vert {\bf x}_i,0)$ is then a genuine transition probability, which  greatly simplifies the calculation  of $Z_A(\lambda,t)$ by avoiding the lengthy computation of the path integral over ${\bf X}$. Introducing the $\lambda$-dependent friction coefficient
\begin{align}
\label{Eqgammaeff}
\widetilde \gamma(\lambda)=[(\gamma +\gamma')^2-4\lambda \gamma \gamma']^{1/2} \ ,
\end{align}
we indeed obtain
\begin{align}
\label{EqSlambdaM1}
{\cal S}_{\lambda}[{\bf X}]&\equiv \frac{1}{4\gamma} \int_0^t dt'\:[m\dot v_{t'}+\widetilde \gamma(\lambda) v_{t'} +\bar k x_{t'}]^2\nonumber\\
& +\frac{\gamma +\gamma' -\widetilde \gamma(\lambda)}{4\gamma} [\bar k (x_f^2-x_i^2)+m(v_f^2-v_i^2)]\ ,
\end{align}
and the time-extensive part of this action is the OM functional corresponding to the effective Langevin equation
\begin{align}
\label{EqLeff}
m\dot v_t= -\widetilde \gamma(\lambda) v_t-\bar k x_t+\sqrt{2\gamma T}\xi_t \ .
\end{align}
Eq. (\ref{EqZAexact}) then becomes
\begin{align}
\label{EqZwexact1}
Z_A(\lambda,t)&=e^{\frac{\gamma +\gamma' -\widetilde \gamma(\lambda)}{2m}t} \int d{\bf x}_i\: p({\bf x}_i) \int d{\bf x}_f f_{A,\lambda}({\bf x}_i,{\bf x}_f)\nonumber\\
&\times e^{-\beta \frac{\gamma +\gamma' -\widetilde \gamma(\lambda)}{4\gamma} [\bar k (x_f^2-x_i^2)+m(v_f^2-v_i^2)]}p_{\tilde \gamma}({\bf x}_f,t\vert {\bf x}_i,0)\ ,
\end{align}
where $p_{\tilde \gamma}({\bf x}_f,t\vert {\bf x}_i,0)$ is the transition probability associated with the dynamics  (\ref{EqLeff}). (The extra exponential factor $e^{-\frac{\widetilde \gamma(\lambda)}{2m}t}$ in Eq. (\ref{EqZwexact1}) comes from the contribution of the effective friction coefficient $\widetilde \gamma(\lambda)$ to the Jacobian.) 
Since $\widetilde \gamma(0)=\gamma +\gamma'$ and $f_{A,\lambda=0}({\bf x}_i,{\bf x}_f)=1$, it is readily seen that  $Z_A(0,t)$ is properly normalized. 

To proceed further, we replace $p({\bf x})$ by its expression in the stationary state~\cite{MR2012} 
\begin{align}
\label{Eqpst}
p({\bf x})=\frac{\beta\sqrt{\bar k m}}{2\pi}\frac{\gamma +\gamma'}{\gamma}e^{-\beta\frac{\gamma +\gamma'}{2\gamma }[\bar k x^2+mv^2]}\ ,
\end{align}
and we compute $p_{\tilde \gamma}({\bf x}_f,t\vert {\bf x}_i,0)$  by using the relation $p_{\tilde \gamma}({\bf x}_f,t\vert {\bf x}_i,0)=p_{\tilde \gamma}({\bf x}_f,t;{\bf x}_i,0)/p_{\tilde \gamma}({\bf x}_i)$. The pdf $p_{\tilde \gamma}({\bf x}_i)$ is given by Eq. (\ref{Eqpst}) with  $\gamma+\gamma'$ replaced by $\widetilde \gamma(\lambda)$, and $p_{\tilde \gamma}({\bf x}_f,t;{\bf x}_i,0)$ is given by the standard formula for the joint probability density of a  two-dimensional  Ornstein-Uhlenbeck process~\cite{R1989},
\begin{align}
\label{Eqptilde}
p_{\tilde \gamma}({\bf x}_f,t;{\bf x}_i,0)=\frac{1}{4\pi^2\sqrt{\mbox{det}{\bf \Phi}}} e^{-\frac{1}{2} {\bf B}^T{\bf \Phi}^{-1}{\bf B}}\ ,
\end{align}
where 
\[
{\bf \Phi}(\lambda,t)=\left(
\begin{array}{cccc}
   \phi_{xx}(0,\lambda)&0 & \phi_{xx}(t,\lambda) &\phi_{xv}(t,\lambda) \\
  0&\phi_{vv}(0,\lambda) &\phi_{xv}(-t,\lambda) &\phi_{vv}(t,\lambda)    \\
  \phi_{xx}(t,\lambda)&\phi_{xv}(-t,\lambda) & \phi_{xx}(0,\lambda)& 0   \\
  \phi_{xv}(t,\lambda)&\phi_{vv}(t,\lambda) & 0 & \phi_{vv}(0,\lambda)
\end{array}
\right) \ ,
\]
and
\[
\label{EqvectorB}
{\bf B}\equiv\left(
\begin{array}{c}
  x_i   \\
  v_i   \\
  x_f \\
  v_f    
\end{array} 
\right) 
\]
is the $4$-dimensional vector representing the initial and final conditions. The functions $\phi_{xx}(t,\lambda),\phi_{xv}(t,\lambda)$ and $ \phi_{vv}(t,\lambda)$ are the stationary time-dependent correlation functions associated with Eq. (\ref{EqLeff}) (see Ref.~\cite{MR2012} for the full expressions). In particular, $\phi_{xx}(0,\lambda)=\gamma T/(\widetilde \gamma(\lambda)\bar k)$ and $\phi_{vv}(0,\lambda)=\gamma T/(\widetilde \gamma(\lambda)m)$. Plugging all these expressions into  Eq. (\ref{EqZwexact1}) and carrying out the Gaussian integrals over $x_i,v_i$ and $x_f,v_f$, we finally obtain the compact result
\begin{align}
\label{EqZAlambda}
Z_A(\lambda,t)=\frac{1}{\sqrt{\mbox{det}({\bf 1}+{\bf \Phi}{\bf L}_A)}}\frac{\gamma+\gamma'}{\widetilde\gamma(\lambda)} e^{\mu (\lambda)t} \ ,
\end{align}
where  
\begin{align}
\label{Eqmuwlim}
\mu(\lambda)=\frac{1}{2m}[\gamma + \gamma'-\widetilde \gamma(\lambda)]\ ,
 \end{align}
and
  \[
{\bf L}_A(\lambda)=\frac{1}{\gamma T}\left(
\begin{array}{cccc}
 \bar k h_A^+(\lambda)&0 &0 & 0 \\
  0&mh_A^+(\lambda)& 0 & 0  \\
 0&0 & \bar kh_A^-(\lambda)&0   \\
 0&0 &0& mh_A^-(\lambda) 
\end{array}
\right) \ ,
\]
with 
\begin{align}
\label{EqhA}
&h_W^{\pm}(\lambda)=m\mu(\lambda)\nonumber\\
&h_Q^{\pm}(\lambda)=m\mu(\lambda)\pm \lambda \gamma\nonumber\\
&h_{\Sigma}^{\pm}(\lambda)=m\mu(\lambda)\mp \lambda \gamma'\ .
 \end{align}
It turns out that $\Sigma_t=\beta {\cal W}_t+(\gamma'/\gamma) \beta \Delta {\cal U}$ in the stationary state, which explains that $h_{\Sigma}^{\pm}(\lambda)$ is obtained from $h_Q^{\pm}(\lambda)$ by  interchanging $\gamma$ and $\gamma'$ and flipping the sign of the last term. Note also that the  present definition of $\mu(\lambda)$ differs from that in ~\cite{MR2012}.

$Z_A(\lambda,t)$ is a complicated function of $\lambda$ and the inverse Fourier transform can only be computed numerically. On the other hand, the long-time limit is readily obtained by noting that the  matrix ${\bf \Phi}(\lambda,t)$  becomes diagonal when $t\rightarrow \infty$, provided  $\lambda <\lambda_{max}=(\gamma+\gamma')^2/(4\gamma \gamma')$ so that  $\widetilde \gamma(\lambda)$ and thus $\mu(\lambda)$ are real~\cite{MR2012}. Then
\begin{align}
\lim_{t\rightarrow \infty} \sqrt{\mbox{det}({\bf 1}+{\bf \Phi}{\bf L}_A)}=\frac{[\widetilde \gamma(\lambda)+h_A^+(\lambda)][\widetilde \gamma(\lambda)+h_A^-(\lambda)]}{\widetilde\gamma(\lambda)^2}\ ,
\end{align}
which leads to
\begin{align}
\label{EqZwasymp}
Z_A(\lambda,t)\sim \frac{(\gamma +\gamma')\widetilde \gamma(\lambda)}{[\widetilde \gamma(\lambda)+h_A^+(\lambda)][\widetilde \gamma(\lambda)+h_A^-(\lambda)]}e^{\mu(\lambda)t}\ .
\end{align}
We can thus identify the three different prefactors as 
\begin{subequations}
\label{EqgA:subeqns}
\begin{align}
g_W(\lambda)&=\frac{4(\gamma +\gamma')\widetilde \gamma(\lambda)}{[\gamma +\gamma'+\widetilde \gamma(\lambda)]^2}\label{EqgA:subeq1}\\
g_Q(\lambda)&=\frac{4(\gamma +\gamma')\widetilde \gamma(\lambda)}{[\gamma +\gamma'+\widetilde \gamma(\lambda)]^2-4\lambda^2\gamma^2}\label{EqgA:subeq2}\\
g_{\Sigma}(\lambda)&=\frac{4(\gamma +\gamma')\widetilde \gamma(\lambda)}{[\gamma +\gamma'+\widetilde \gamma(\lambda)]^2-4\lambda^2\gamma'^2}\label{EqgA:subeq3}\ .
\end{align}  
\end{subequations} 

One can check that Eq. (\ref{Eqmuwlim}) is also given by the general expression (\ref{Eqmuw1}) of $\mu(\lambda)$ in the small-$\tau$ limit. Indeed, the response function $\chi(\omega)$ associated with Eq. (\ref{EqMR}) reads
\begin{align}
\chi(\omega)= [-m\omega^2-i(\gamma+\gamma')\omega+\bar k]^{-1}\ ,
\end{align}
and the function $H_{\lambda}(\omega)$ in Eq. (\ref{Eqmuw1}) (in the original dimensionfull units) is now given by 
\begin{align}
H_{\lambda}(\omega)^{-1}&= \vert\chi(\omega)\vert ^{-2}-4 \lambda\gamma \gamma'\omega^2 \nonumber\\
&= [\bar k-m\omega^2]^2+\omega^2[(\gamma+\gamma')^2-4 \lambda \gamma \gamma'] \ .
\end{align}
This can be identified for {\it all} values of $\lambda<\lambda_{max}$ with the square modulus of the   response function associated with  the effective Langevin Eq. (\ref{EqLeff}),
\begin{align}
\label{Eqchilambda}
\chi_{\lambda}(\omega)\equiv  (-m\omega^2-i\widetilde \gamma(\lambda)\omega +\bar k)^{-1}\ .
\end{align}
Eq. (\ref{Eqmuw1}) then reads
\begin{align}
\label{Eqmulim}
\mu(\lambda)&=\frac{1}{2\pi}\int_{-\infty}^{+\infty}d\omega \:\ln \frac{\chi_{\lambda}(\omega)}{\chi_0(\omega)}\ ,
\end{align}
where we have eliminated the modulus since the imaginary part of $\chi(\omega,\lambda)$ is an odd function of $\omega$. The two poles of $\chi_{\lambda}(\omega)$ lie on the lower-half of the complex $\omega$-plane for all values of the parameters, and by using a contour similar to the one considered in Fig. 4 of I (replacing $\omega$ by $is$), one recovers Eq. (\ref{Eqmuwlim}) from Cauchy's residue theorem. As it must be, Eq. (\ref{Eqmuwlim}) also agrees with the expression of the SCGF obtained in ~\cite{Sab2011} when the temperature $T'$ of the second thermostat is set to zero (this is also true for $g_Q(\lambda)$ given by Eq. (\ref{EqgA:subeq2})). Interestingly,  $\mu(\lambda)$ and the three prefactors $g_A(\lambda)$ are independent of the spring constant $\bar k$.

As discussed in Sec. IV.B.1, the value $\lambda=1$ deserves special attention. From Eq. (\ref{Eqgammaeff}), one obtains $\widetilde \gamma(1)=\vert \gamma-\gamma'\vert$, so that  Eqs. (\ref{Eqmuwlim}) and (\ref{Eqchilambda}) yield 
\begin{align}
\label{EqAmu1}
\mu(1)=\begin{cases} \frac{\gamma'}{m} \ , \ \mbox{for} \ \gamma\ge \gamma'\\ \frac{\gamma}{m} \ ,\  \mbox{for }\ \gamma'\ge \gamma \end{cases}
\end{align}  
and 
\begin{align}
\label{Eqchi1}
\chi_1(\omega)=\begin{cases} [-m\omega^2-i(\gamma-\gamma')\omega +\bar k]^{-1}\equiv \widetilde \chi(\omega) \ , \ \mbox{for} \ \gamma\ge \gamma'\\ [-m\omega^2-i(\gamma'-\gamma)\omega +\bar k]^{-1}\equiv\hat \chi(\omega) \ ,\  \mbox{for }\ \gamma'\ge \gamma\ ,\end{cases}
\end{align} 
where $\widetilde \chi(\omega)$ and $\hat \chi(\omega)=\widetilde \chi(-\omega)$ are the response functions obtained from the transformations $\gamma \to -\gamma$ and $\gamma' \to -\gamma'$, respectively, which correspond to the so-called ``hat" and ``tilde" conjugate dynamics defined in the main text (changing $\tau$ into $-\tau$ in the small-$\tau$ limit is indeed equivalent to flipping the sign of $\gamma'$). 

\subsection{Fluctuations of the work}

We now use Eq. (\ref{EqZwasymp}) to investigate how the work ${\cal W}_t[{\bf X}]=-\gamma'\int_0^t dt'\: v_{t'}^2$ fluctuates in the long-time limit. The key point is that the prefactor $g_W(\lambda)$ defined by Eq. (\ref{EqgA:subeq1}) has no singularity, so that the LDF $I(w)$ is always given by the Legendre transform of $\mu(\lambda)$,  with the saddle point $\lambda^*(w)$  solution of the equation
\begin{align}
\frac{1}{m}\frac{\gamma \gamma'}{\widetilde \gamma(\lambda^*)}+w=0\ .
\end{align}
This yields
\begin{align}
\lambda^*(w)=\frac{\gamma'}{4\gamma}[\frac{(\gamma+\gamma')^2}{\gamma'^2}-\frac{\gamma^2}{m^2w^2}]
\end{align}
for $w<0$, whereas there is no solution for $w>0$. The function $\lambda^*(w)$ decreases monotonically from $\lambda_{max}$ to $-\infty$ as $w$ increases from  $-\infty$ to $0$, 
and the LDF  is then given by
\begin{align}
I(w)=-\frac{(\gamma+\gamma')^2}{\gamma\gamma'}\frac{(w-\bar w)^2}{4w}\  \  \mbox{for} \ w<0 \ ,
\end{align}
where $\bar w=-\gamma\gamma'/[m(\gamma +\gamma')]$ is the average work rate. (Note that the LDF is defined here  with the same sign as in Ref.~\cite{T2009}, whereas the opposite convention was adopted in Ref.~\cite{MR2012}.)

We next  consider the long-time behavior of $Z_W(1,t)=\langle e^{-\beta {\cal W}_t[{\bf X}]}\rangle$ to point out a mistake in Ref.~\cite{KQ2007}. According to Eq. (18) in that paper,  one should have the asymptotic fluctuation relation  $\lim_{t \to \infty} \langle e^{-\beta {\cal W}_t[{\bf X}]+\Delta S_{pu}(t)}\rangle=1$, where $\Delta S_{pu}(t)$ is the so-called ``entropy pumping" contribution, which is equal to $-(\gamma'/m)t$  in the present model (as the feedback force depends linearly on the velocity). On the other hand, the exact asymptotic expression  (\ref{EqZwasymp}) yields  $\lim_{t \to \infty} \langle e^{-\beta {\cal W}_t[{\bf X}]-\mu(1)t}\rangle=g_W(1)$, which is a different result. First, $\mu(1)$ is equal to $\gamma'/m$  for $\gamma\ge \gamma'$ only (cf. Eq. (\ref{EqAmu1})). Second, Eq. (\ref{EqgA:subeq1}) states that $g_W(1)=1-(\gamma'/\gamma)^2 $  for $\gamma\ge \gamma'$ and $g_W(1)= 1-(\gamma/\gamma')^2 $  for $\gamma'\ge \gamma$. In both cases, this is different from $1$. The error  in Ref.~\cite{KQ2007} consists in assuming that $\beta {\cal W}_t[{\bf X}]$ always fluctuates like $\Sigma_t[{\bf X}]$ asymptotically because the two observables  only differ by a temporal boundary term. However, this term may have large fluctuations of  order $t$, as discussed below.

\subsection{Fluctuations of the dissipated heat and the entropy production}

We now turn our attention to $Z_Q(\lambda,t)$ and $Z_{\Sigma}(\lambda,t)$. We first notice from Eqs. (\ref{EqZAlambda})-(\ref{EqhA}) that the two generating functions are related to one another by interchanging $\gamma$ and $\gamma'$, a symmetry that is not obvious from the mere definition of the observables. Although the long-time behavior of $Z_{\Sigma}(\lambda,t)$ has already been investigated in Ref.~\cite{MR2012}, it is worth revisiting this analysis to stress some important points that were left aside.

We  know from Eq. (\ref{EqIFTQ}) in the main text that the heat ${\cal Q}_t[{\bf X}]$ satisfies at all times the  IFT 
 \begin{align}
\label{EqIFTQ1}
\langle e^{-\beta  {\cal Q}_t}\rangle=e^{\frac{\gamma}{m}t}\ .
\end{align} 
The symmetry $\gamma\leftrightarrow \gamma'$ thus implies that 
 \begin{align}
\label{EqIFTQ2}
\langle e^{-\Sigma_t}\rangle=e^{\frac{\gamma'}{m}t}\ ,
\end{align}  
which is  the IFT obtained in Ref.~\cite{KQ2007} and re-derived  in Ref.~\cite{MR2012}. In the long-time limit, these two relations imply  that $\mu_Q(1)=\gamma/m$ and $\mu_{\Sigma}(1)=\gamma'/m$. Comparing with Eq. (\ref{EqAmu1}) we thus see that  $\mu(1)$ differs from $\mu_Q(1)$ for $\gamma>\gamma'$ and from $\mu_{\Sigma}(1)$ for $\gamma'>\gamma$.
There is no contradiction, however, and the mismatch can be ascribed to  rare but large fluctuations of the temporal boundary terms that are not included in the definition (\ref{EqZAlambda}) of $\mu(\lambda)$ (and more generally in the calculation that leads to Eq. (\ref{Eqmuw1}) in the main text). As is clear from Eqs. (\ref{EqgA:subeq2}) and (\ref{EqgA:subeq3}), the mathematical consequence is the divergence of the prefactors $g_Q(1)$ for $\gamma\ge \gamma'$ and  $g_{\Sigma}(1)$ for $\gamma'\ge \gamma$. 
\begin{figure}[hbt]
\begin{center}
\includegraphics[width=8cm]{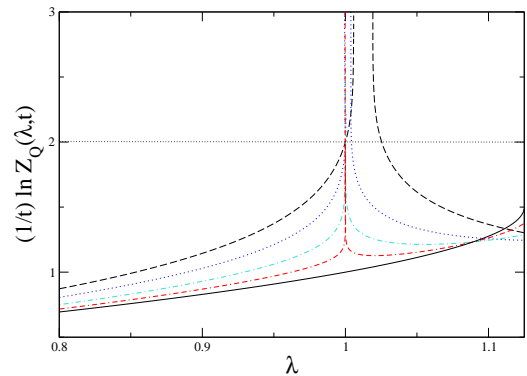}
 \caption{\label{FigB1} Behavior of $(1/t)\ln Z_Q(\lambda,t)$ as a function of  $\lambda$ in the vicinity of $\lambda=1$  for $\gamma=2$ and $ \gamma'=1$ ($m=1,\bar k=1,T=1$). From top to bottom: $t=3,5,10,25$. Observe that $(1/t)\ln Z_Q(1,t)=\gamma/m=2$ for all values of $t$. The  solid black line shows the theoretical SCGF $\mu(\lambda)$ given by Eq. (\ref{Eqmuwlim}).}
\end{center}
\end{figure}

To understand more precisely what is going on, let us investigate the behavior of $Z_Q(\lambda,t)$  for finite  $t$.  (Of course, the same analysis holds for $Z_{\Sigma}(\lambda,t)$ by changing $\gamma$ into $\gamma'$.) The key observation is that the determinant of  the matrix ${\bf 1}+{\bf \Phi}\:{\bf L}_Q$  in Eq. (\ref{EqZAlambda}) vanishes at $\lambda=\lambda_+(t)>1$ and that this zero moves towards $1$ as $t \rightarrow \infty$. The determinant is negative beyond  this value but  becomes positive again for larger values of $\lambda$.  The resulting behavior of $(1/t)\ln Z_Q(\lambda,t)$ is illustrated in Fig. \ref{FigB1}. Note that the intermediate region where the determinant is negative and $Z_Q(\lambda,t)$ imaginary shrinks as $t$ increases. As it must be, one has  $(1/t)\ln Z_Q(1,t)=\gamma/m$ at all times.

A careful analysis of Eq. (\ref{EqZAlambda}) shows that the  behavior of  $Z_Q(\lambda,t)$ for $t$ large but finite and $\lambda$  close to $1$ is described by the boundary-layer expression
\begin{align}
\label{EqBL}
Z_Q(\lambda,t) \sim \frac{(\gamma+\gamma')(\gamma-\gamma')^2e^{t\gamma'/m}}{2\gamma^3 \vert 1-\lambda\vert \sqrt{1+B(u,t)}},
\end{align}
where 
\begin{align}
\label{EqBL}
B(u,t)&=\frac{(\gamma+\gamma')(\gamma-\gamma')^2}{\gamma^3}\frac{4km-(\gamma-\gamma')^2 \cos(\alpha t/m)}{\alpha^2}u\nonumber\\
&+\frac{(\gamma+\gamma')^2(\gamma-\gamma')^4}{4\gamma^6}u^2\ ,
\end{align}
with the scaling variable $u=(1-\lambda)^{-1}e^{-t(\gamma-\gamma')/m}$  and $\alpha=\sqrt{4\bar km-(\gamma-\gamma')^2}$ (which is here assumed  to be real). Accordingly, one has 
\begin{align}
Z_Q(\lambda,t)\sim \frac{(\gamma+\gamma')(\gamma-\gamma')^2}{2\gamma^3 \vert 1-\lambda\vert }e^{\frac{\gamma'}{m}t}
\end{align}
for $u\ll 1$, i.e., $t/\tau_0\gg\frac{\gamma}{\gamma-\gamma'}\ln \frac{1}{\vert 1-\lambda\vert}$ (where $\tau_0=m/\gamma$ is the viscous relaxation time for $\gamma'=0$), and 
\begin{align}
Z_Q(\lambda,t)\sim e^{\frac{\gamma}{m}t}
\end{align}
for $u\gg 1$, i.e., $t/\tau_0\ll \frac{\gamma}{\gamma-\gamma'}\ln \frac{1}{\vert 1-\lambda\vert}$. This crossover behavior, which is reminiscent of a smoothed dynamical first-order transition, is illustrated in Fig. {\ref{FigB2}. \begin{figure}[hbt]
\begin{center}
\includegraphics[width=8cm]{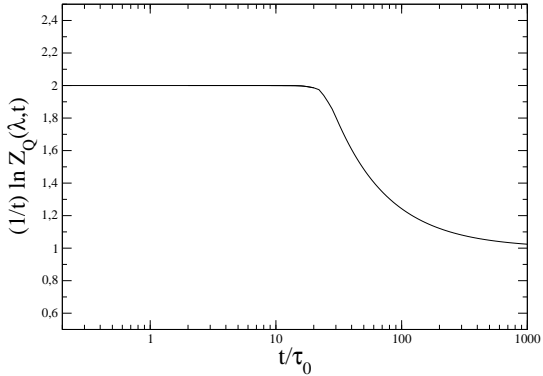}
 \caption{\label{FigB2} Behavior of $(1/t)\ln Z_Q(\lambda,t)$ as a function of  $t/\tau_0$ for $\gamma=2,\gamma'=1$ and $1-\lambda=10^{-6}$ ($m=1,\bar k=1,T=1$). Note the crossover from $\gamma$ to $\gamma'$ around $t/\tau_0=\frac{\gamma}{\gamma-\gamma'}\ln \frac{1}{1-\lambda}\approx 28$. The crossover time decreases as  $\lambda$ moves away from $1$.}
\end{center}
\end{figure}

As it turns out, $g_Q(\lambda)$ has also another pole at $\lambda_-=-(1+2\gamma'/\gamma)$, which in contrast with the pole at $\lambda=1$ exists for both $\gamma\ge \gamma'$ and $\gamma' \ge \gamma$. More generally, for $t$ finite, $Z_Q(\lambda,t)$ diverges at $\lambda=\lambda_-(t)<\lambda_-$. This singularity moves towards $\lambda_-$ as $t$ increases, and is equal to $\lambda_-$ at a {\it finite} critical time $t_c$. (Alternatively, when regarded  as a function of $t$, $Z_Q(\lambda,t)$ diverges  at a certain time $t(\lambda)\le t_c$ for $\lambda\le \lambda_-$.) The behavior of $Z_Q(\lambda,t)$ in the vicinity of $\lambda_-$ is thus different from the behavior in the vicinity of $\lambda=1$. On the other hand, this kind of behavior is observed in other nonequilibrium models, see e.g. \cite{KNP2011}.

Two comments are in order:

1)  The two poles of $g_Q(\lambda)$ have a different origin, as can be seen by performing  the averages over the initial and final conditions  in Eq. (\ref{EqZwexact1}) separately. The pole at $\lambda=1$ for $\gamma\ge \gamma'$ comes from the average over ${\bf x}_f$, whereas the pole at $\lambda=\lambda_-$ comes from the average over ${\bf x}_i$. This can also be seen by
taking  the long-time limit directly in Eq. (\ref{EqZwexact1}) using the fact that $p_{\tilde \gamma}({\bf x}_f,t\vert {\bf x}_f,0)\rightarrow p_{st,\tilde \gamma}({\bf x}_f)$ as $t\rightarrow\infty$. 

2) These poles are {\it not} the poles of $g_{\Delta U}(\lambda)$. Indeed, a simple calculation shows that the generating function of  $\Delta {\cal U}$ behaves asymptotically as
\begin{align}
\label{EqZUexact}
Z_{\Delta U}(\lambda,t)\sim \frac{(\gamma +\gamma')^2}{(\gamma +\gamma')^2-\gamma^2\lambda^2}\ .
\end{align}
Its domain of definition is thus $[-\frac{\gamma+\gamma'}{\gamma},\frac{\gamma+\gamma'}{\gamma}]$, which is not the domain of definition of $Z_{\cal Q}(\lambda,t)$.  This results from the fact that the boundary term in Eq. (\ref{EqZwexact1}) (for ${\cal A}_t=\beta{\cal Q}_t$) does not only comes from the function  $f_{Q,\lambda}=e^{\lambda \beta \Delta {\cal U}({\bf x}_i,{\bf x}_f)}$. In other words, ${\cal W}_t$ and $\Delta {\cal U}$ cannot be treated as uncorrelated random variables asymptotically, as is often assumed~\cite{BJMS2006,TC2007,TC2009,GSPT2013}. 
As a consequence, the slope of the LDF $I(q)$, which is determined by the poles of $g_Q(\lambda)$ in a certain range of $q$, is not related to the tails of the pdf of $\Delta {\cal U}$. Explicitly, we find

a) For $\gamma' >\gamma$,

\begin{equation}
\label{Eqasym1}
I(q)=\left \{\begin{aligned}
& -\frac{(\gamma+\gamma')^2}{\gamma\gamma'}\frac{(q-\bar q)^2}{4q} \quad \mbox{for $q \le q_1$}\\       
&\frac{\gamma'}{m}+(1+2\frac{\gamma'}{\gamma})q \quad \mbox{for $q \ge q_1$} \\
\end{aligned}
 \right.
\end{equation}
where $\bar q=-\gamma\gamma'/[m(\gamma +\gamma')]$ and $q_1=-\gamma \gamma'/[m(\gamma +3\gamma')]$ (such that $\lambda^*(q_1)=\lambda_-$).

b) For $\gamma>\gamma'$,
\begin{equation}
\label{Eqasym1}
I(q)=\left \{\begin{aligned}
&-\frac{\gamma'}{m}-q & \quad \mbox{for $q \le q_2$}\\       
& -\frac{(\gamma+\gamma')^2}{\gamma\gamma'}\frac{(q-\bar q)^2}{4q}& \quad \mbox{for $q_2\le q \le q_1$} \\
&\frac{\gamma'}{m}+(1+2\frac{\gamma'}{\gamma})q& \quad \mbox{for $q \ge q_1$} \\
\end{aligned}
 \right.
\end{equation}
where $q_2=-\gamma\gamma'/[m(\gamma -\gamma')]$ (such that $\lambda^*(q_2)=1$).


\begin{thebibliography}{10}
\bibitem{note35} We could as well replace $\widetilde \chi(\omega)$ in Eq. (\ref{Eqmuw2}) by $\hat \chi(\omega)=\widetilde \chi(-\omega)$, the response function of the $\gamma \to -\gamma$  ``hat" dynamics, since $H_{1}(\omega)$ is an even function of $\omega$. However, we prefer to use $\widetilde \chi(\omega)$ in order to directly exploit the analysis performed in I.
\bibitem{note40} Take for instance the derivative with respect to $g$, close the integration contour by a large semi-circle on the r.h.s. and use Jordan's lemma for integrals of the type $\int dz e^{iz}f(z)$.
In passing, note that causality is not  a sufficient condition for the integral to vanish. The conditions of Jordan's lemma must also be satisfied. For instance, if one expands $\chi(s)$ at first order in $\tau$, one has $\lim_{\vert z\vert\rightarrow \infty} \vert z f(z)\vert\ne 0$ and the integral is then equal to $g\tau/(2Q_0)$.
\bibitem{note45} Case (c) occurs for instance for $Q_0=2$, $g/Q_0=0.55$ and $\tau= 8.5$, which corresponds to the second stability lobe in Fig. 2 of I. Then $\mu(1)\approx 0.040$ and $\dot S_{\cal J}\approx 0.075$, whereas $1/Q_0=1/2$. Note that the feedback control heats the system for these values of the parameters, as can be seen in Figs. 10 and 16 of I.
\bibitem{note56} The cusp in $\dot S_{\cal J}$ results from the intricate evolution of the poles of $\widetilde \chi(s)$ with $\tau$, as explained in I. This nonanalytical behavior is rounded in the numerical estimate of $\mu_{\Sigma}(1)$, presumably because of finite-time effects.
\bibitem{note3} In particular, for $\tau=7.6$ and $\lambda=1$, we find $g_W(1)\approx 0.95$, in very good agreement with the prediction of Eq. (\ref{Eqgw10}).
\bibitem{note159} The calculation, which involves expanding all quantities, including path probabilities, up to second order in  the feedback gain, is long and very technical, and we do not give the details here. They are available  upon request.
\end{thebibliography}
\end{document}